\documentclass[a4paper,fleqn,usenatbib]{mnras}

\usepackage[T1]{fontenc}
\usepackage{ae,aecompl}

\usepackage{graphicx}	
\usepackage{amsmath}	
\usepackage{amssymb}	

\title[Supra-galactic colour patterns in GCS]{Supra-galactic colour patterns in globular cluster systems}

\author[J.C. Forte]{
Juan C. Forte,$^{1,2}$\thanks{E-mail: planeta.jcf@gmail.com}
\\
$^{1}$Consejo Nacional de Investigaciones Cient\'ificas y T\'ecnicas (CONICET), Argentina\\
$^{2}$Planetario `Galileo Galilei', Ministerio de Modernizaci\'on, Innovaci\'on y Tecnlog\'ia, Ciudad Aut\'onoma de Buenos Aires, Argentina
}
\date{Accepted:March 11, 2017; Received: March 5, 2017; in original form: January 29, 2017}
\pubyear{2017}
\begin{document}
\label{firstpage}
\pagerange{\pageref{firstpage}--\pageref{lastpage}}
\maketitle
\begin{abstract}
   An analysis of globular cluster systems associated with  galaxies 
  included in the Virgo and Fornax $HST$-$Advanced~Camera~Surveys$
  reveals distinct $(g-z)$ colour modulation patterns. These features 
  appear on composite samples of globular clusters and, most evidently,
  in galaxies with absolute magnitudes $M_{g}$ in the range from -20.2 to 
  -19.2. These colour modulations are also detectable on some samples
  of globular clusters in the central galaxies $NGC~1399$  and $NGC~4486$ (and confirmed
  on data sets obtained with different instruments and photometric systems),
  as well as in other bright galaxies in these clusters.
  After discarding field contamination, photometric errors and statistical 
  effects, we conclude that these supra-galactic colour patterns
  are real and reflect some previously unknown characteristic. These
  features suggest that the globular cluster formation process was
  not entirely stochastic but included a fraction of clusters that
  formed in a rather synchronized fashion over large spatial scales,
  and in a tentative time lapse of about 1.5 $Gy$ at redshifts $z$ between 2 and 4.
  We speculate that the putative mechanism leading to that synchronism may be
  associated with large scale feedback effects connected with violent
  star forming events and/or with super massive black holes. 
\end{abstract}
%
%
%
%
%
%
%

\section{Introduction}
\label{Intro}
%
Globular clusters ($GCs$), both as individuals and as members of
systems ($GCS$) in different galaxies, can be genuinely considered as crucial and
ever-surprising Rosetta stones. The astrophysical information they contain,
 however, has not been completely decoded yet.
Among other issues, the presence of multiple stellar populations
 \citep[e.g.][]{Milone2017} within a given cluster is a
relatively new challenge that adds to other historical questions: How 
do they form and why this process is not evident in low $z$ galaxies,
except in particular cases (e.g. galaxy mergers), where some
objects seem to resemble the young counterparts of $GCs$.

A prominent characteristic of $GCS$ in massive galaxies is that their
globular cluster colour distributions ($GCCDs$ in what follows) are frequently
"bimodal", i.e., these distributions are dominated by two ("blue" and "red")
 cluster families. This feature is  known since the mid 90's and an account
of the contributions that led to its discovery is given in \citet{Brodie2006}.    
As shown by \citet{Peng2006}, the presence of "bimodality" depends on
the  absolute magnitude of a given galaxy. Blue globulars are present in
practically all galaxies while the red $GC$ sub-population becomes more 
important as the galaxy brightness increases.

In general, blue $GCs$ exhibit shallow spatial distributions and seem connected
 with dark matter halos. In turn, red $GCs$ show more concentrated distributions, that are
 comparable with the galaxy surface brightness, at least in the inner regions 
 \citep{Faifer2011, Forbes2012, Forbes2016, Forte2012}.
Recent works \citep*{Dabrusco2015} indicate the presence of complex
structures on top of those overall spatial patterns.

In a broad sense, bimodality seems just to reflect the halo-bulge 
structures common to most massive galaxies.
The description of $GC$ populations in terms of two components, however, is
probably an over simplification of the general situation. 
An updated and thorough summary of the arguments that justify an effort
 to advance "beyond bimodality" is given by  \citet{Harris2017}.

This work presents an exploratory attempt in that direction. The underlying issue
 is if, besides the usual decoding of the $GCCDs$ in terms of age, chemical abundance,
 and spatial distribution, there might be other (more subtle) features,
 providing a different kind of information in these distributions.

An interesting aspect, in this last context, is the possible
detectability of relics of energetic events on the stellar
populations in the early Universe that, later on, survived until today. As
frequently stated, globular clusters  appear as remarkable
survivors born in those ages  \citep[e.g.][]{Forbes2015, Vanzella2016}
and  could be a suitable workbench to  test that possibility.

Observational evidence in the sense that powerful nuclear outflows
may have been common, leaving imprints not only in the host galaxy but
also on larger spatial scales, is reviewed in \citet{Fabian2012}. In
this scenario, nuclear super massive black holes ($SMBH$) may have
played a significant role.
 
The importance of both star formation and $AGNs$
 on the parameters that characterize the interstellar medium, and 
 have an impact on that process, has become clear in recent years as
 shown by galaxy formation models \citep[e.g][]{Vogelsberger2014}.

The literature includes some previous works that have discussed a possible connection
 between high energy phenomena and $GCs$. In this frame, the remarkable homogeneity of
 the (almost) ever present blue (halo) $GCs$, suggests that their formation 
 may have been regulated by large spatial scale events like, for example,
 the re-ionization of the Universe \citep[e.g.][]{Cen2001}.

The usual procedure to study the $GCCDs$ along the years has been 
based on discrete  and/or "generalized" (smoothed) histogram versions
of these distributions.

In this paper we present a somewhat different approach. Instead of 
analysing $GCs$ as members of a single galaxy, we work on composite 
samples of clusters that belong to galaxies within a given sampling window, 
defined in the  galaxy absolute magnitude $M_{g}$ $vs.$ $(g-z)$ colour
plane. 

This procedure will presumably erase the  "personality" of a given $GCS$
but  might enhance the presence of systemic features, if any, common 
to these galaxies. For example, $GC$ colour patterns that could arise
as a result of old feedback phenomena.

The core of the analysis is the $ACS$ photometry of $GCs$ associated with 
 galaxies in the the Virgo and Fornax clusters presented, respectively,
 by \citet{Jordan2009} and \citet{Jordan2015}.
In the case of the central and massive galaxies $NGC~1399$ and $NGC~4486$,
 we revisit the $GCs$ photometry given in \citet*{Forte2007} and \citet{Forte2013}.

The structure of the paper is as follows: Section 2 gives a summary 
of the characteristics of the photometric data, assumed distance 
moduli and interstellar reddening; Section 3 presents an overall view
of the $GC$ colour distributions for galaxies in Virgo and Fornax, both
in the form of discrete and generalized histograms;
the procedure adopted for the recognition of eventual patterns in the
$GC$ colour distributions is described in Section 4; in Section 5, we
present the  analysis of $GCs$ in Virgo and Fornax galaxies with $M_{g}=$
-20.2 to -19.2; Section 6 describes an attempt to identify the colour patterns
 in $GCs$ associated with the central galaxies  $NGC~1399$ and $NGC~4486$; a
 tentative decomposition of the $GCCD$ in  terms of discrete mono-colour 
populations is described in Section 7; Finally, Sections 8 and 9
present a discussion of the results and the conclusions,
respectively.
\section{Data sources}
\label{sec2}
The details concerning the $GC$ photometry in the Virgo and
Fornax $ACSs$ have been extensively discussed in \citet{Jordan2009} and
 \citet{Jordan2015}. The galaxy morphology in these surveys is dominated by
 elliptical and lenticular galaxies.

In what follows, the $GC$ $(g-z)$ colours are corrected for interstellar
reddening and taken from Table 4 and Table 2, for Virgo and Fornax $GCs$,
respectively, in those papers. Aiming at an homogeneous comparison between the
 $GCS$ in Virgo and Fornax,  we adopted upper galactocentric radii of 110 $\arcsec$ and 90.6
$\arcsec$, respectively. These radii correspond to a galactocentric radius of
  8.8 $Kp$c after adopting the $SBF$ distance moduli given by \citet{Blakeslee2009}.

We also set an apparent magnitude $g=$ 25.0 (about a magnitude
above the $ACS$ limiting magnitude) as the faintest limit of the analysis.
This limit guarantees both a high areal completeness level and mean colour 
errors of $\approx$ 0.07 mag.

\begin{figure}
	\includegraphics[width=\columnwidth]{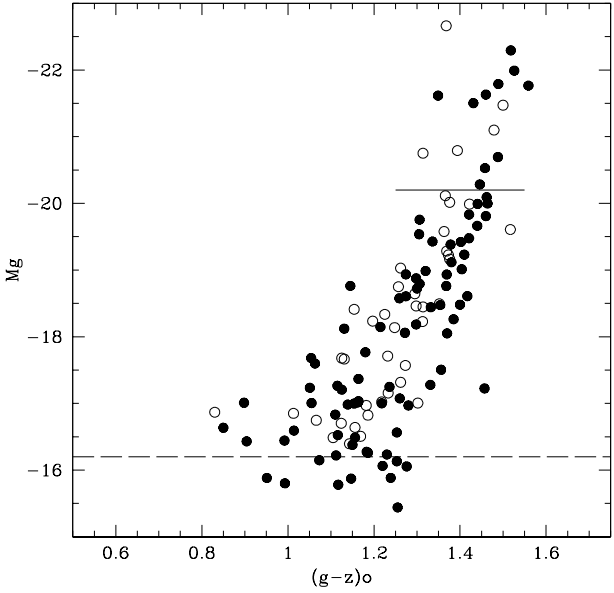}
    \caption{Absolute $M_{g}$ magnitude $vs.$ (g-z)o colours for Virgo
 (filled circles) and Fornax (open circles) $ACS$ galaxies. Galaxies brighter than 
 $M_{g}$=-20.2 (solid line) are considered as "giants". The dashed line at
  $M_{g}$=-16.2 defines the faint boundary of the analysis.   
}
    \label{fig:fig1}
\end{figure}

For old $GCs$ (i.e. 12 $Gy$) we expect a $(g-z)o$ colour range from
0.75 to 1.55  (see, for example, \citealt{Forte2013}). In this work we
adopt a wider colour window ($(g-z)o$ from 0.50 to 1.80), aiming at including
eventually younger or older clusters. The absolute $M_{g}$ magnitude vs $(g-z)$ colour
 diagram for 88 galaxies in Virgo and 42 galaxies in Fornax is displayed in 
Fig.~\ref{fig:fig1}

The  $M_{g}$ magnitudes were derived from the $B$ band photometry by
\citet*{Binggeli1984} and \citet*{Ferguson1989}, the relation between $g$ and
$B$ magnitudes given by \citet*{Chonis2008}, and the $(g-r)$ vs $(g-z)$
relation for galaxy colours we derived from \citet{Chen2010}.

Both the  galaxy magnitudes and colours were corrected adopting the
reddening maps presented in \citet*{Schlafly2011}.
From these maps, the mean colour excess for Virgo is $E_{(B-V)}=$0.03
$\pm$ 0.01 and $E_{(B-V)}=$0.01 $\pm$ 0.005 for Fornax. The formal $rms$
errors of both values are very low but, in fact, the $rms$ of the original
infra-red emission $vs.$ colour excess map calibration is considerably larger,
and close to  $\pm$ 0.03 mag.

Giant galaxies lie above the horizontal line at $M_{g}=$-20.2 in 
 Fig.~\ref{fig:fig1}. For these objects the areal coverage
 of the $ACSs$ is relatively low (see  Fig.4 in \citealt{Forte2014}).
 Some of these objects, given their status as central and dominant
 galaxies, are discussed in Section 6.

 For the particular analysis of the central galaxies $NGC~4486$ and $NGC~1399$, 
 we re-analyse the photometry presented by \citet{Forte2007}. These homogeneous
 ground based observations were obtained with the the $Mayall$ and $Blanco$ 4-m 
 telescopes at $KPNO$ and $CTIO$, adopting the $(C-T_{1})$ colour index defined 
 with pass bands of the Washington photometric system.
 In the case of $NGC~4486$, we also revisit the $Gemini-GMOS$ photometry for
 some 500 $GCs$ located in a peripheral field ($\approx$ 2.5 $\arcmin$ to the south
 of the galaxy  centre).
\section{Discrete and smoothed GCCD in Virgo and Fornax galaxies}
\label{sec3}
The usual procedure in the analysis of the $GCCD$ is the
adoption of discrete colour bin histograms. The size of these 
bins is commonly set on an assessment of the colour errors (e.g. adopting
a bin size comparable to the photometric errors), or aiming at
decreasing the statistical noise of the  data through variable bin size 
histograms, as in \citet{Peng2006}.

An alternative approach are smoothed colour distributions 
that can be derived by convolving the colour data with, for example, Gaussian kernels
whose standard deviation parameters are also set in terms of the photometric errors.\\
Needless to say that this smoothing decreases the resolution in colour even more, as
 it adds to the "blurring" already introduced by photometric errors.
      
In both cases (i.e, discrete bins or smoothed data), the central issue is the compromise between 
colour resolution and the signal to statistical noise ratio.
In the search for systemic features, however, the smoothed $GCCDs$ are more sensitive
to data "clumpines" and allow the identification of subtle structures, provided
the adoption of a proper value for the smoothing kernel.

In this paper we adopt both representations of the $GCCDs$. The bin size of discrete 
histograms, and the Gaussian kernel parameter ($\sigma_{(g-z)}$) were tuned
 up in such a way that, the eventual peaks seen both in the discrete bins
 and in the smoothed data, do not differ by more than 0.02 mags.

Experimenting with these parameters on the $ACSs$ photometric data, we find that, 
 0.04 mag bins and a $\sigma_{(g-z)}$=0.015 mag Gaussian kernel, meet that condition.
Smoothing kernels larger than $\approx$ 0.04 mag., practically erase all the
substructure present in the $GCCDs$ and just leave, depending the case, broad uni modal
or bimodal distributions. 

\begin{figure}
	\includegraphics[width=\columnwidth]{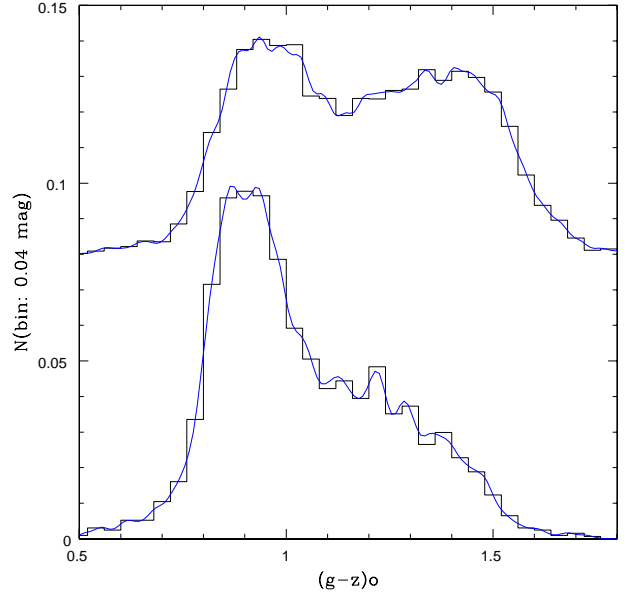}
    \caption{Composite (g-z)o colour distributions for 4506 GCs brighter 
than $M_{g}$=-6.5 in 10 giant galaxies (upper histogram;
shifted upwards arbitrarily) and 3165 $GCs$ in 68 galaxies (fainter than $M_{g}$=-20.2;
 lower histogram) in the Virgo cluster. The histograms have 0.04 mag bins. The solid 
curves are smoothed colour distributions after convolving with a 0.015 mag Gaussian
kernel. A broad bimodality is seen for $GCs$ in giant galaxies and becomes less
evident in fainter galaxies. 
}
    \label{fig:fig2}
\end{figure}

\begin{figure}
	\includegraphics[width=\columnwidth]{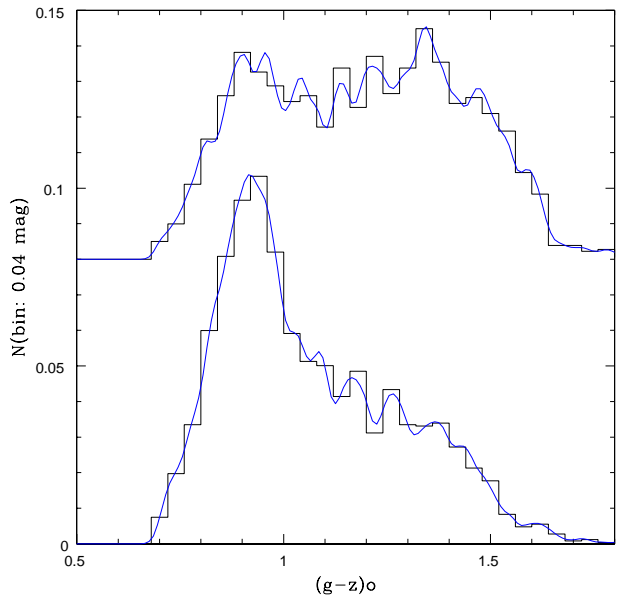}
    \caption{Composite (g-z)o colour distributions for 1789 $GCs$  brighter 
than $M_{g}$=-6.5 in 5 giant galaxies (upper histogram;
arbitrarily shifted upwards) and 2528 $GCs$ in 37 galaxies (fainter than $M_{g}$=-20.2;
(lower histogram) in the Fornax cluster. The histograms have 0.04 mag bins. The solid 
curves are smoothed  colour distributions after convolving with a 0.015 mag Gaussian 
kernel. A broad bimodality is seen for $GCs$ in giant galaxies and becomes less evident
in fainter galaxies. 
}
    \label{fig:fig3}
\end{figure}

Fig.~\ref{fig:fig2} and Fig.~\ref{fig:fig3} display the composite $GCCD$, in
both formats, for $GCs$ brighter than $M_{g}=-6.5$, in giant galaxies 
(upper curves), and galaxies in the absolute magnitude range from $M_{g}$=-20.2
and $M_{g}$=-16.2 (lower curves).

The sampled $GC$ populations for galaxies brighter than $M_{g}$=-16.2 amounts
to 7671 $GCs$ in 88 Virgo galaxies and 4317 clusters in 42 Fornax
galaxies. In these populations, 4506 clusters belong to ten giant galaxies (i.e.
above the $M_{g}$=-20.2 line in Fig. 1) in Virgo and 1789 clusters in five Fornax
galaxies.

Both Fig.~\ref{fig:fig2} and Fig.~\ref{fig:fig3} show a result already found
and emphasized by \citet{Peng2006}: $GC$ colour bimodality 
becomes more evident as the galaxy brightness increases.
These authors, and \citet{Jordan2009}, present a description  of their method to
segregate $GCs$ from field interlopers using the $z$ magnitude $vs.$ cluster size 
diagram (see their Fig. 1).

 At our adopted limiting magnitude ($g$=25.0),
and considering the expected  colour range for $GCs$, the limiting $z$
magnitude results close to 24.50. Figure 1 in \citet{Peng2006} indicates
that, above this magnitude, the presence of field interlopers is practically
negligible (also see Section 8).
\section{Colour pattern recognition}
\label{sec4}
The recognition of eventual patterns arising in feedback phenomena on the $GCSs$
faces a starting problem: The a priori unknown characteristics of these 
patterns. In a naive way, one might expect that quenching or enhancement
of $GCs$ formation would be reflected, respectively, as valleys and peaks 
on top of the overall cluster formation history, encoded in the $GCCD$, of a given galaxy.
Here we adopt a simple strategy based  on the identification of peaks and
 valleys in the  composite $GCCDs$.

Both the Virgo and Fornax galaxies were sampled with a moving window, characterized
 by a range in galaxy absolute magnitudes $\Delta M_{g}$. This windows moves in steps
 $\delta_{M_{g}}$. Discrete and smoothed colour distributions were then
obtained by convolving the $(g-z)o$ colours of all $GCs$ associated with
galaxies in the sampling window with a Gaussian kernel $\sigma_{(g-z)}$.
Peaks and valleys in the composite (smoothed) $GCCDs$ where 
identified  at colours were the first derivative $dN/d(g-z)o$ is null.

After exploring the effects of changing the $\Delta M_{g}$, $\sigma_{(g-z)})$
and $\delta_{M_{g}}$ values, in what follows we adopt values of 0.40, 0.015 and 0.20, 
respectively, for these parameters.

\begin{figure}
	\includegraphics[width=\columnwidth]{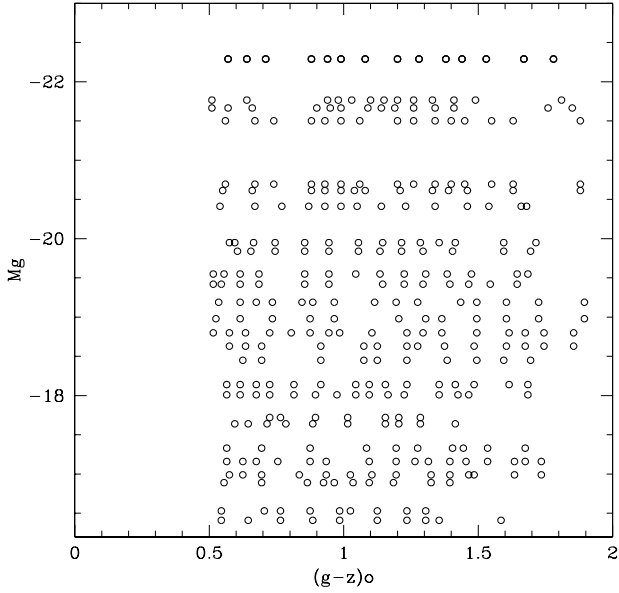}
    \caption{Colour peaks (open circles), detected on composite $GC$ samples
for Virgo galaxies, grouped in 0.4 mag bins in $M_{g}$ . These colours are
shown as a function of the mean composite $M_{g}$ of the galaxies each group.
}
    \label{fig:fig4}
\end{figure}

\begin{figure}
	\includegraphics[width=\columnwidth]{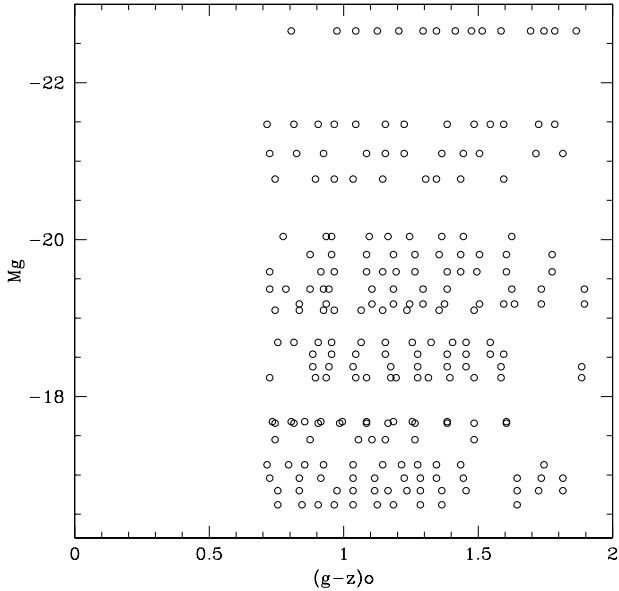}
    \caption{Colour peaks (open circles), detected on composite $GC$ samples for
Fornax galaxies, grouped in 0.4 mag bins in $M_{g}$. These colours are shown
as a function of the mean composite magnitude $M_{g}$ of the galaxies in 
each group.
}
    \label{fig:fig5}
\end{figure}

The colour peaks detected within each sampling window are shown, as a
function of the mean composite $M_{g}$ magnitude of the galaxies, in 
Fig.~\ref{fig:fig4} and Fig.~\ref{fig:fig5} for the Virgo and Fornax 
$GCs$ respectively.

\begin{figure}
	\includegraphics[width=\columnwidth]{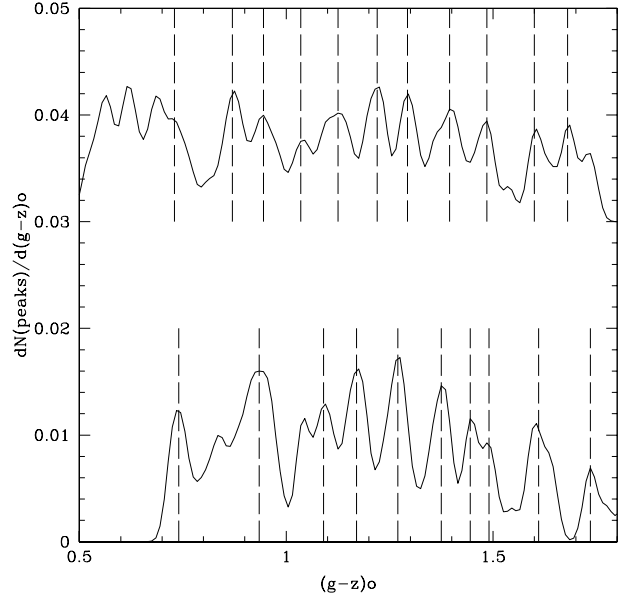}
    \caption{Smoothed distributions of the colour peaks (shown in Fig. 4 and
Fig. 5) after a convolution with a 0.015 mag Gaussian kernel. The upper curve
corresponds to $GCs$ in non-giant Virgo galaxies (arbitrarily shifted upwards) and the lower
 one to $GCs$ in non-giant Fornax galaxies. Dashed lines identify the colour of each peak in
 the different patterns. 
}
    \label{fig:fig6}
\end{figure}

As a first step we analyse galaxies in the magnitude range from $M_{g}$=-20.2
to -16.2 (i.e., non-giant galaxies).
 The number $vs.$ colour statistics for these peaks in Virgo (237 peaks; 
upper curve), and Fornax (179 peaks; lower curve), are shown in 
Fig.~\ref{fig:fig6} in a smoothed format, normalized by the total number of peaks,
 and adopting the same Gaussian kernel $\sigma_{(g-z)}$, used to derive the
 $GCCDs$ within the galaxy sampling windows.

A simple cross correlation of these patterns with the composite $GCCDs$
defined by galaxies within a given $\Delta M_{g}$ window, helps in
identifying which galaxy groups are the most important contributors
to the presence of these patterns.

It is worth mentioning that the colour patterns and the $GCCDs$ are two different
 types of statistics. The first reflects the frequency of a given $(g-z)$ 
 colour peak for galaxy groups within a determined range in absolute magnitude.
 The second, indicates the  frequency of $GCs$ with the same colour.\\
This procedure indicates that the highest correlation signals are obtained, 
both in Virgo and Fornax, for $GCs$ in galaxies within the five brightest
 sampling windows, i.e., $M_{g}$ from -20.2 to -18.8. 

Within those windows, the routine finds a total of 80 colour peaks in
Virgo and 64 in Fornax. In what follows we refer to the smoothed colour 
distribution of these peaks as the $template$ Virgo and Fornax patterns
($TVP$ and $TFP$) respectively.

\begin{figure}
	\includegraphics[width=\columnwidth]{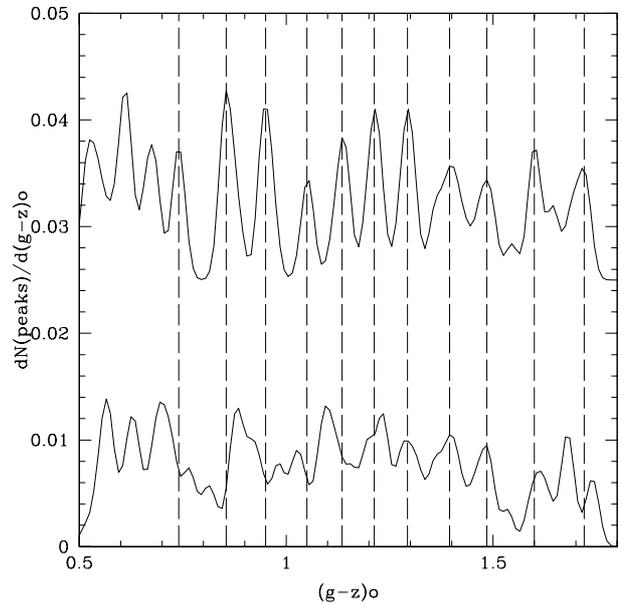}
    \caption{Smoothed colour distribution for 80 colour peaks detected in $GCs$
associated with non-giant Virgo galaxies  brighter than  $M_{g}=$-18.8 (upper curve; shifted
arbitrarily upwards) compared with those corresponding to 141 peaks found in fainter
 galaxies (lower curve). The dashed lines indicate the colours corresponding to peaks
in the the upper curve.
}
    \label{fig:fig7}
\end{figure}

\begin{figure}
	\includegraphics[width=\columnwidth]{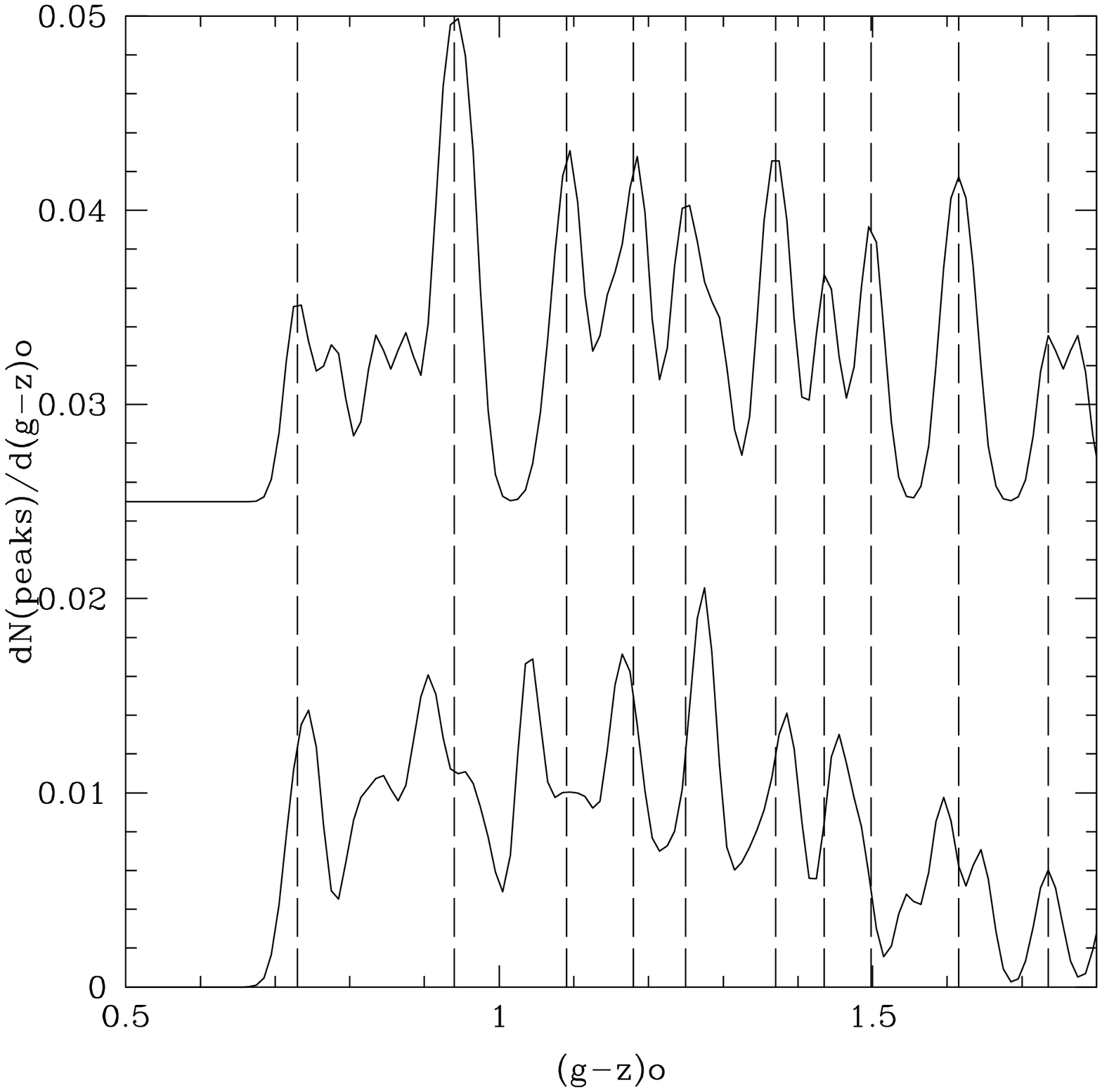}
    \caption{Smoothed colour distribution for 64 colour peaks detected in $GCs$
associated with non-giant Fornax galaxies brighter than $M_{g}=$-18.8 (upper curve; arbitrarily shifted
upwards) compared with those corresponding to 115 peaks found in all the fainter galaxies
 (lower curve). The dashed lines indicate the peak colours corresponding to the upper curve.
}
    \label{fig:fig8}
\end{figure}

The template colour patterns are listed in Table~\ref{table_1}.
 From here, the references to a particular colour peak is given within
brackets an identified by their number (first column) in this table. Dubious
 cases are indicated by a colon following the identification number.
\begin{table}
\centering
\caption{Template (g-z) colour patterns in Virgo and Fornax.}
\begin{tabular}{c c c c c}
\hline
\hline
\textbf{Id. Number}& &\textbf{(g-z)o~Virgo}& &\textbf{(g-z)o~Fornax}\\
\hline
 1& &0.74& &0.73\\
 2& &0.85& &0.94\\
 3& &0.95& &1.09\\
 4& &1.05& &1.18\\
 5& &1.13& &1.25\\
 6& &1.21& &1.37\\
 7& &1.29& &1.43\\
 8& &1.39& &1.50\\
 9& &1.48& &1.61\\
10& &1.60& &1.73\\
11& &1.72& &----\\
\hline
\label{table_1}
\end{tabular}
\end{table}
The $TVP$ and $TFP$ derived from  galaxy groups between $M_{g}=$ -20.2 to -18.8, are 
compared with those corresponding to $GCs$ in galaxy groups fainter than $M_{g}$=-18.8
 (159 peaks in Virgo and 115 peaks in Fornax) in  Fig.~\ref{fig:fig7} and Fig.~\ref{fig:fig8}.

 For Virgo, a comparison of both colour patterns shows coincidences in the colour
 domain of the intermediate-red $GCs$: peaks [6,7,8,9,10], are comparable to within
 $\pm 0.01$ mags in $(g-z)$. Bluer peaks, in turn, do not show a clear correspondence.
 
\begin{figure}
	\includegraphics[width=\columnwidth]{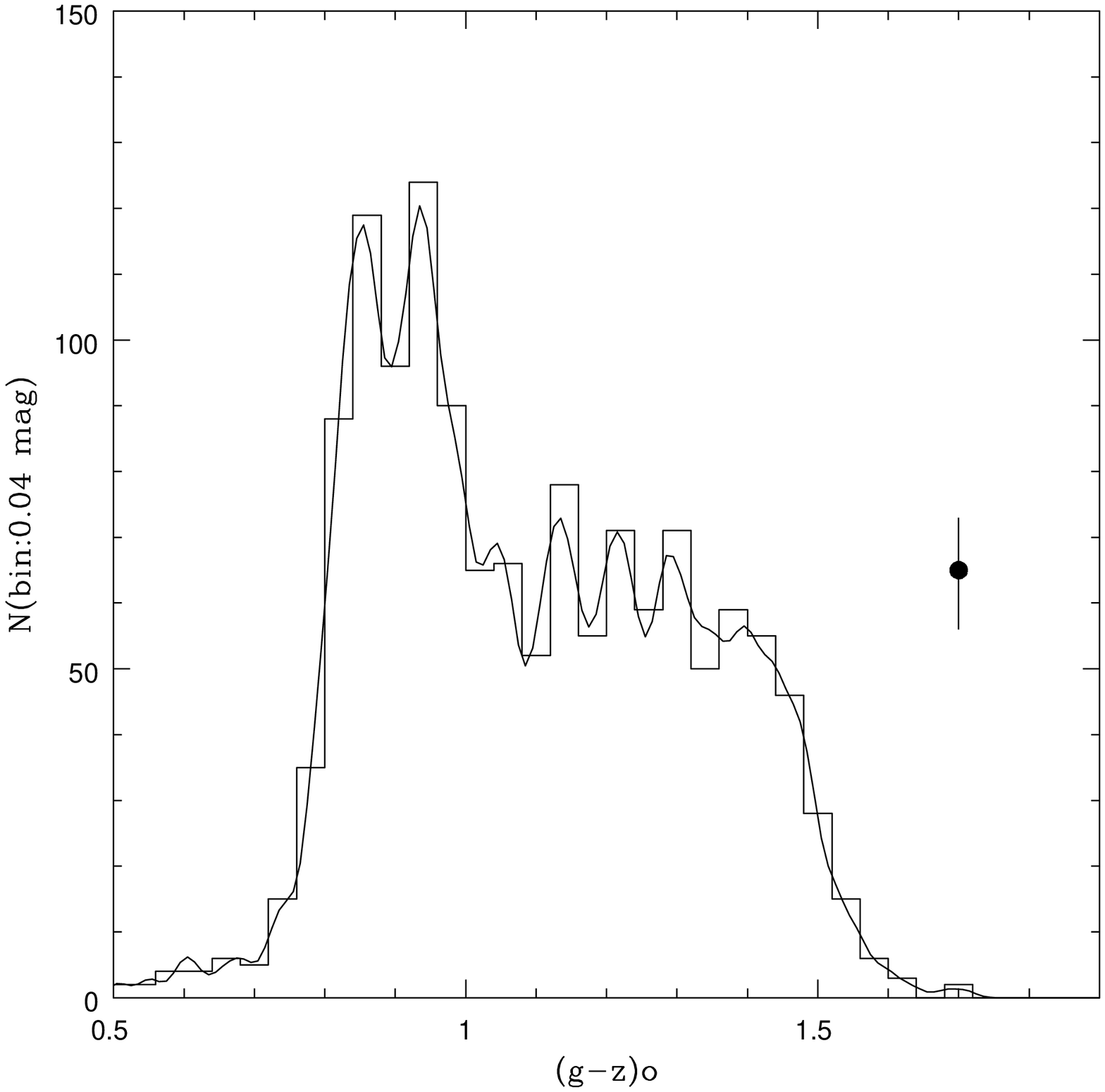}
    \caption{Discrete and smoothed (g-z)o colour distribution for  clusters
 brighter than $M_{g}$$\approx$-6.4. The sample includes 1349 $GCs$ in 13 Virgo galaxies with $M_{g}$
 between -20.2 and -19.2. The histogram has 0.04 mag bins while the smoothed distribution 
 corresponds to a 0.015 mag Gaussian kernel. The filled circle at right corresponds to a $GC$ 
 level of 65 clusters per bin. The vertical line indicates the expected counting uncertainty.
}
    \label{fig:fig9}
\end{figure}

\begin{figure}
	\includegraphics[width=\columnwidth]{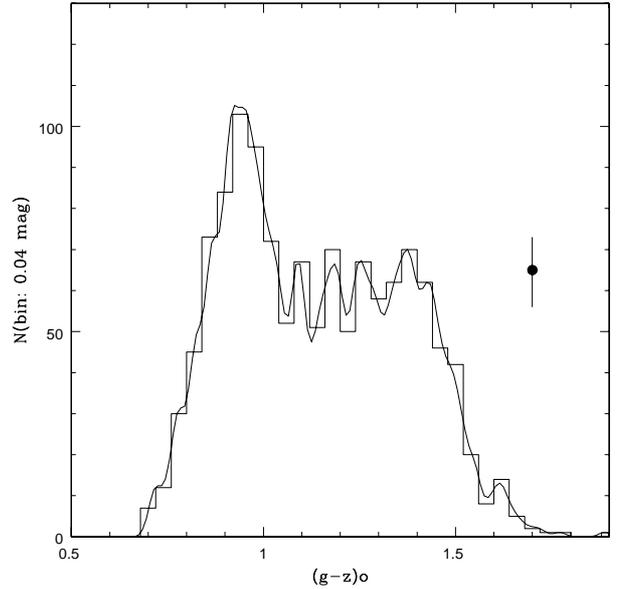}
    \caption{Discrete and smoothed (g-z)o colour distribution for clusters
 brighter than $M_{g}$$\approx$-6.4. the sample includes 1266 $GCs$ in 7 Fornax galaxies with $M_{g}$
 between -20.2 and -19.2. The histogram has 0.04 mag bins, while the smoothed distribution 
 corresponds to a 0.015 mag Gaussian kernel. The filled circle at right corresponds to a $GC$ level
 of 65 clusters per bin. The vertical line indicates the expected counting uncertainty.
}
    \label{fig:fig10}
\end{figure}

 The Fornax patterns also show coincidences [4,5,6,7,8], to within $\pm$ 0.025 mag, in the
 colour range of the intermediate-red $GCs$. The weaker coincidences, compared to those
 in Virgo, are possibly a consequence of the smaller number statistics in Fornax.
 Again, as in the case of the Virgo patterns, the coincidences are not detectable for
 peaks bluer than $(g-z)$=$\approx$ 1.10. 
The discrete and smoothed $GCCDs$ for clusters brighter than $M_{g}$$\approx$-6.4 ($g=$25
at the distance of the Fornax cluster) in galaxies within the absolute magnitude
 range $M_{g}$ from -20.2 to -19.2, are displayed in Fig.~\ref{fig:fig9} (13 Virgo
 galaxies; 1349 $GCs$) and in Fig.~\ref{fig:fig10} (7 Fornax galaxies, 1266 $GCs$). 
These figures also show the expected  statistical count uncertainties
 for a population level of $\approx$ 65 $GC$ per bin (shown as bar crossed dots).

$GCs$ in these Virgo galaxies exhibit a double blue peak [i.e., clusters 
bluer than $(g-z)o=$1.0] and five colour peaks in the domain of the intermediate-red 
$GCs$. In turn, the $GCCD$ in Fornax has a single blue peak and
four in the intermediate-red $GCs$ colour range. In both cases, the amplitude of the
fluctuations are comparable to the expected statistical uncertainties.
\section{GC colour distributions in non-giant galaxies.}
\label{sec5}
In this section we analyse the behaviour of the $GCCDs$
in galaxies fainter than $M_{g}$=-20.2, and leave the
analysis of some relevant giant galaxies for a following Section.
All the $GCCDs$ presented in what follows are normalized
by the total number of clusters in each sample.

\subsection{GCs in Virgo Galaxies.}
\label{subsec5a}
\subsubsection{GCCD for galaxies with $M_{g}=$-20.2 to -19.2}
\label{sss51}
\begin{figure}
	\includegraphics[width=\columnwidth]{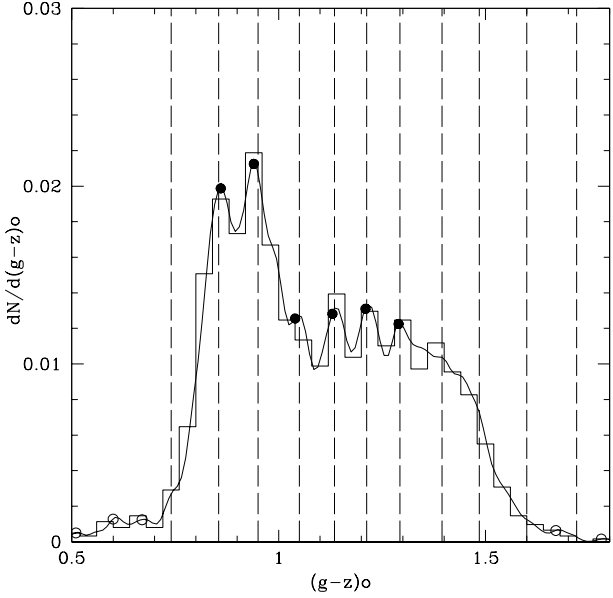}
    \caption{Smoothed (g-z)o colour distribution (adopting a 0.015 mag Gaussian kernel)
 corresponding to 1513 $GCs$ in 13 Virgo galaxies with $M_{g}$ -20.2 to -19.2 compared
 with the $TVP$ colours (dashed lines; see text).
}
    \label{fig:fig11}
\end{figure}

The composite $GCCD$ corresponding to thirteen galaxies includes 1513 $GCs$ brighter than
$g=$25.0 and is displayed in Fig.~ \ref{fig:fig11}. Dashed lines in this,
and in the following diagrams of this subsection, correspond to
the $TVP$. Filled dots indicate the colour peaks, found by the routine
described in Section 4. The routine also computes the difference between the colour
 of each peak  and the nearest colour peak in the corresponding template pattern. An
 overall $rms$  of the differences is obtained taking into account all
 the detected peaks. Then, a small colour shift is allowed (for the whole set of
 detected peaks) in order to minimize that $rms$. This shift may be justified, for example,
 in terms of the uncertainties in interstellar reddening. Alternatively, 
 the shift may reflect a difference in chemical abundance between $GCs$ in
 a given galaxy and those in galaxies that define the template patterns.

\begin{figure}
	\includegraphics[width=\columnwidth]{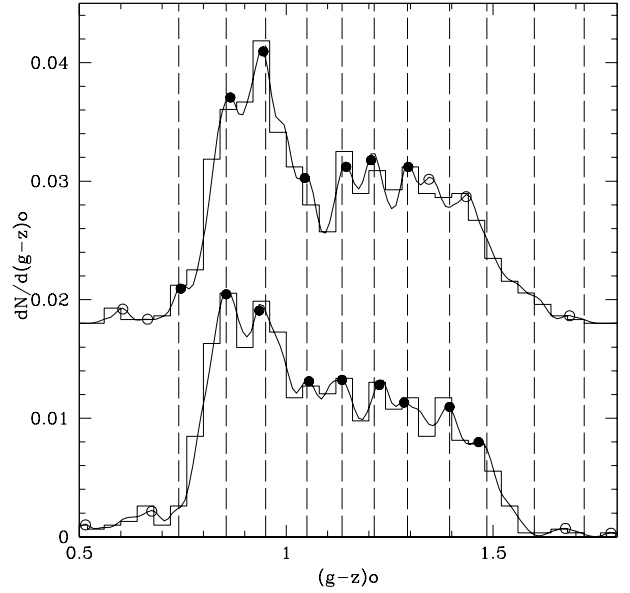}
    \caption{ Smoothed (g-z)o colour distribution (adopting a 0.015 mag Gaussian kernel)
 corresponding to 773 $GCs$ in 6 Virgo galaxies with $M_{g}$ -20.2 to -19.7 (upper histogram;
 shifted arbitrarily upwards) and 762 $GCs$ in 7 galaxies with $M_{g}=$-19.7 to -19.2 
 (lower curve). The dashed lines corresponds to the $TVP$ colours.
}
    \label{fig:fig12}
\end{figure}

 In the case of the $GC$ sample shown in Fig.~ \ref{fig:fig11}, the $TVP$ shows six coincidences,
 [2,3,4,5,6,7], to within 0.01 mag., with the peaks shown by the $GCCD$, and without requiring 
 a colour shift.
 This result can be expected as most of these galaxies are included in the
 galaxy groups that define the $TVP$. In turn, splitting the galaxy group in
 two sub-groups (brighter and fainter than $M_{g}=$-19.7) leads to the $GCCDs$
 shown in Fig.~\ref {fig:fig12}.

The brighter galaxy group includes six galaxies (VCC 759, 1030,
1062, 1154, 1692, 2092; upper histogram) and 689 $GCs$ while, the
fainter one, has seven galaxies (VCC 369, 944, 1279, 1664, 1720,
 1938, 2000; lower histogram) and an almost identical number of 682 $GCs$.

 Both $GCCDs$ exhibit coincidences with the $TVP$. Peaks [1,2,3,4,5,6,7] are
detected in the brighter galaxy groups while peaks [2,3,4,5,6,7,8,9]
appear on the fainter group. In all these cases the agreement with the $TVP$ 
colours is better than $\pm$ 0.01 mag. (and zero colour shift).

None of these galaxies, except VCC~1664 ($NGC~4564$), show a total coincidence
of their $GCCD$ with the $TVP$ when analysed as individuals. In the case of
 VCC~1664, that contributes with 8 percent of the total $GC$ sample, there are
 eight coincidences [2,3,4,5,6,7,8,9], with colour differences within $\pm$ 0.015 mag.
 The larger colour differences in this galaxy are possibly an statistical effect
 resulting from the smaller size of the $GC$ sample.

The analysis of the whole 1513 $GCs$ sample can be also performed in terms of the $g$ 
 apparent magnitude. For objects brighter than $g=$23.0 the $GCCD$ is dominated by
 the double blue peak while redder $GCs$ (see Fig. A1) exhibit a rather poor coincidence
 with the $TVP$.

 In turn, splitting the $GC$ sample corresponding to objects fainter than $g$=23.0 in two
 equal number samples: 458 clusters with $g=$ 23 to 23.75 and 450 clusters
 with $g$= 23.75 to 24.50 (see Fig. A2) allows the identification of the $TVP$
 peaks [2,3,5,6,7] and [2,3,4,5,6,7,8,9,10] respectively.
 
\subsubsection{GCCD for galaxies with $M_{g}=$-19.2 to -18.2}
\label{sss52}
\begin{figure}
	\includegraphics[width=\columnwidth]{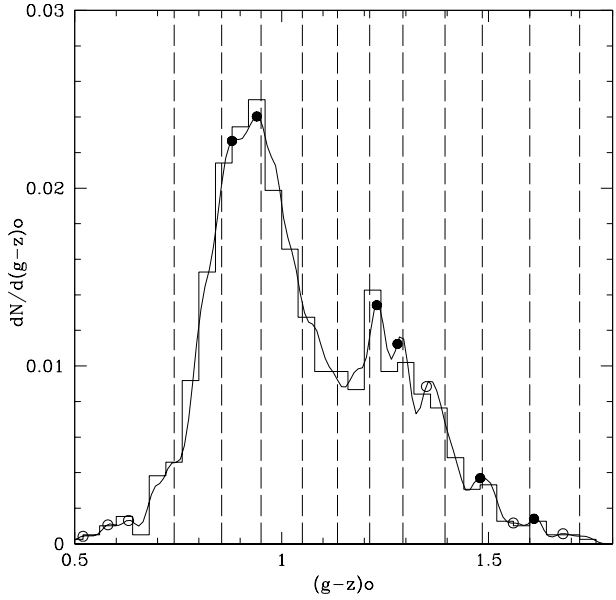}
    \caption{Smoothed (g-z)o colour distribution (adopting a 0.015 mag Gaussian kernel) corresponding to
 883 $GCs$ in 17 Virgo galaxies with $M_{g}$ -19.2 to -18.2 compared with the
 colours of the $TVP$ (dashed lines; see text).
}
    \label{fig:fig13}
\end{figure}

 This group includes 883 $GCs$ in 17 galaxies an the corresponding $GCCD$
 is displayed in Fig.~\ref{fig:fig13}. This distribution shows a broad blue
 peak (two overlapped components ?) and three approximate coincidences
 [6,7,8:] with the $TVP$.
 
\subsubsection{GCCD for galaxies with $M_{g}=$-18.2 to -17.2}
\label{sss53}
\begin{figure}
	\includegraphics[width=\columnwidth]{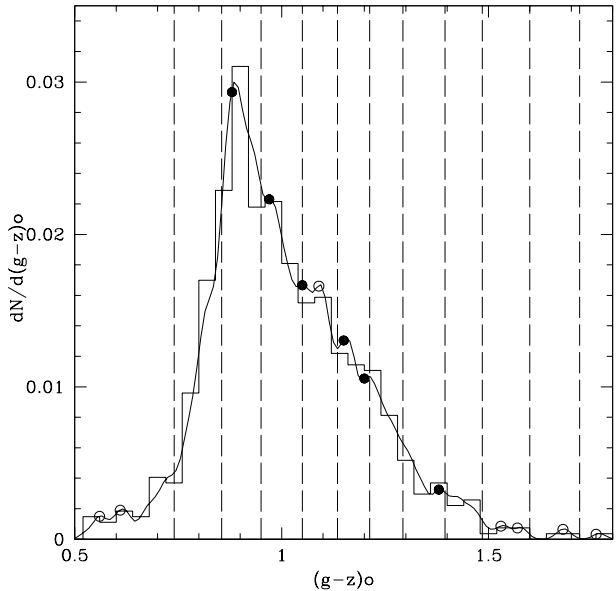}
    \caption{Smoothed (g-z)o colour distribution (adopting a 0.015 mag Gaussian kernel) corresponding to
 673 $GCs$ in sixteen Virgo galaxies with $M_{g}$ -18.2 to -17.2 compared with the $TVP$ colours 
 (dashed lines; see text).
}
    \label{fig:fig14}
\end{figure}

 The $GCCD$ is depicted Fig.~\ref {fig:fig14} and corresponds to 673 $GCs$ in 16 galaxies.
 The blue peak becomes narrower and better defined than in the previous group and
 no further coincidences with the $TVP$ are detectable. This galaxy sample includes
 VCC 1297 ($NGC~4486B$), an object with a relatively large number of $GCs$ for its brightness.
 Possibly, its cluster population includes a fraction of $GCs$ that in fact belong to the nearby
 $NGC~4486$. Removing VCC 1297 from the sampling window, however, does not change
 the shape of the composite $GCCD$ in a significant way. 
\subsubsection{GCCD for galaxies with $M_{g}=$-17.2 to -16.2}
\label{sss53}
\begin{figure}
	\includegraphics[width=\columnwidth]{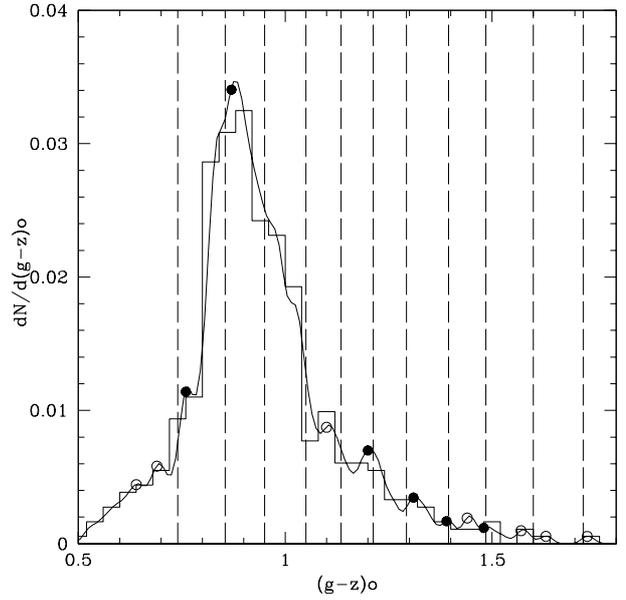}
    \caption{Smoothed (g-z)o colour distribution (adopting a 0.015 mag Gaussian kernel) corresponding to
 453 $GCs$ in 22 Virgo galaxies with $M_{g}$ -17.2 to -16.2 (upper histogram) compared with the
 $TVP$ colours (lower curve; see text).
}
    \label{fig:fig15}
\end{figure}

Fig.~\ref {fig:fig15} depicts the $GCCD$ corresponding to 453 $GCs$ in 22 galaxies.
This $GCs$ sample exhibits a single blue peak, similar to that seen in the previous
 group. Three other $TVP$ peaks [5,6,7] seem to be marginally present.

\subsection{GCs in Fornax galaxies}
\label{subsec5b}
In this subsection, we repeat a similar analysis for $GCs$ in
Fornax galaxies.\\
\subsubsection{GCCD for galaxies with $M_{g}=$-20.2 to -19.2}
\label{sss5b1}
\begin{figure}
	\includegraphics[width=\columnwidth]{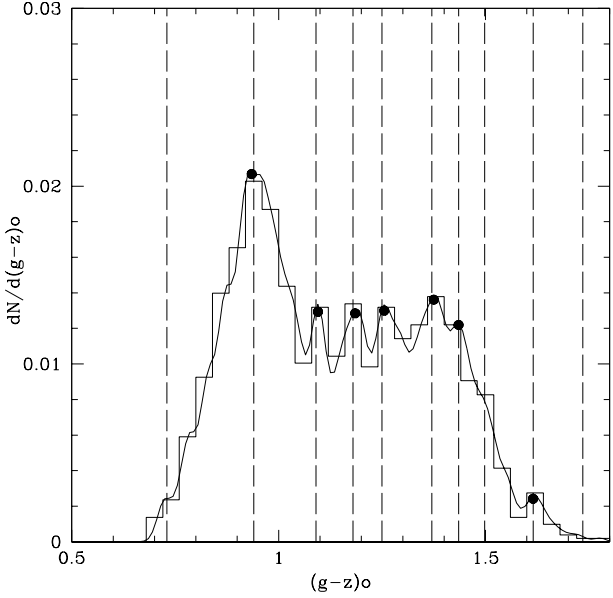}
    \caption{Smoothed (g-z)o colour distribution (adopting a 0.015 mag Gaussian
 kernel) corresponding to 1270 $GCs$ in seven Fornax  galaxies with $M_{g}$ -20.2 to
 -19.2 compared with the colours of the $TFP$  (dashed lines; see text).
}
    \label{fig:fig16}
\end{figure}

The $GCCD$ is displayed in Fig.~\ref{fig:fig16} and includes 1270 $GCs$ in 7 galaxies.
 This distribution shows seven coincidences with the $TFP$, [2,3,4,5,6,7,9]
 within $\pm$ 0.01 or less. As in the case of the Virgo galaxies, this agreement can be
 expected as most of these galaxies define the $TFP$.

\begin{figure}
	\includegraphics[width=\columnwidth]{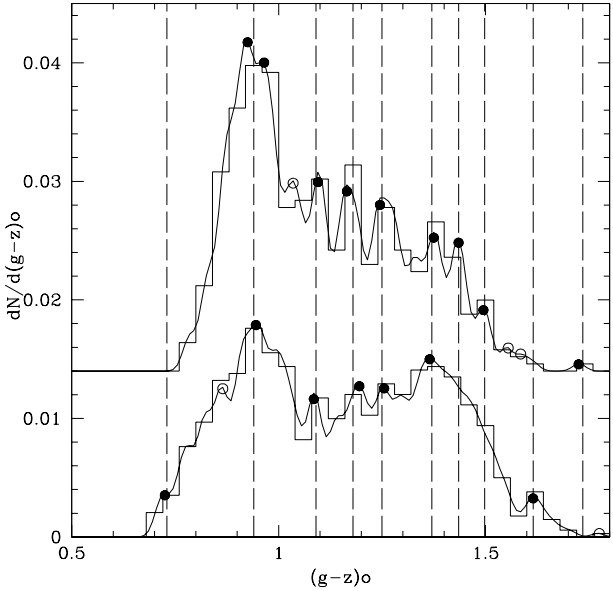}
    \caption{ Smoothed (g-z)o colour distribution (adopting a 0.015 mag Gaussian kernel) corresponding to 594
 $GCs$ in 3 Fornax galaxies with $M_{g}$ -20.2 to -19.7 (upper curve, arbitrarily shifted upwards) and 676 $GCs$ in 4 galaxies with
 $M_{g}$ -19.7 to -19.2 (lower curve). The dashed lines corresponds to the $TFP$ colours (dashed lines).
}
    \label{fig:fig17}
\end{figure}

In turn, the  $GCCD$ for 594 clusters in galaxies in the range
 $M_{g}$=-20.2 to -19.7 (FCC 147, 276, 2006) and for 676 $GCs$ in four galaxies
 in the range  $M_{g}$=-20 to -19.2 (FCC 63, 83, 184, 193) is displayed 
 in  Fig.~\ref{fig:fig17}. In this case the colour distributions indicate eight 
  [2,3,4,5,6,7,8,9] and seven [1,2,3,4,5,6,9] coincidences within $\pm$ 0.01 mags.,
 respectively.

A comparison of the $GCCDs$ for clusters brighter and fainter than $g=$23 in the whole
 sample (1270 clusters; see Fig. A3) shows that the colour pattern is better defined for 
 the bright clusters (in contrast with the results from the Virgo galaxies that have an
 opposite behaviour).

The $GCCDs$ for two sub samples with equal number of clusters (see Fig. A4) show that,
for the fainter clusters, the blue peak is broader and more poorly defined than its
brighter clusters counterparts. For redder $GCs$, the patterns display five coincidences
 in common: [3,4,5,6,7].
As in the case of Virgo, the $TFP$ can be recognized when splitting the $GC$ samples
 in terms of galaxy groups or in terms of apparent $g$ magnitude.  
\subsubsection{GCCD for galaxies with $M_{g}=$-19.2 to -18.2}
\label{sss5b2}
\begin{figure}
	\includegraphics[width=\columnwidth]{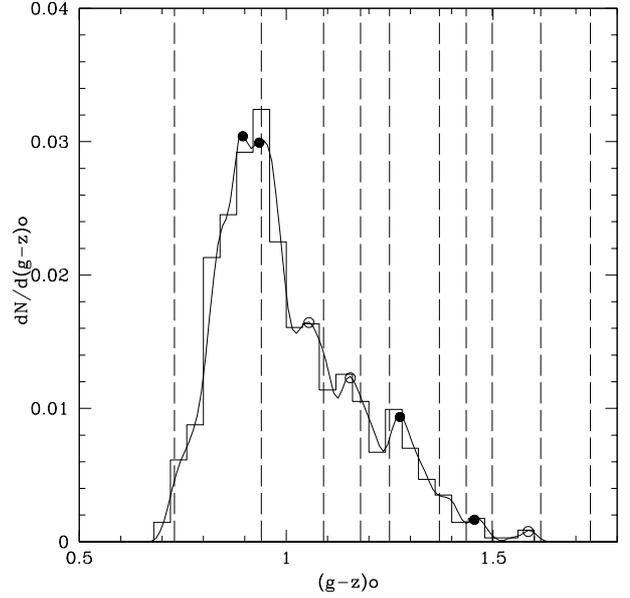}
    \caption{Smoothed (g-z)o colour distribution (adopting a 0.015 mag Gaussian kernel) corresponding to
 856 $GCs$ in 17 Fornax galaxies with $M_{g}$ -19.2 to -18.2 compared with the colours of the $TFP$
 (dashed lines; see text).
}
    \label{fig:fig18}
\end{figure}

Fig.~\ref{fig:fig18} shows the $GCCD$ of 856 $GCs$ in 17 galaxies. The
 blue peak appears located at the expected position of the $TFP$. Three
 mildly modulated features are seen towards the red but they are not clearly
 coincident with the $TVP$.
\subsubsection{GCCD for galaxies with $M_{g}=$-18.2 to -17.2}
\label{sss5b3}
\begin{figure}
	\includegraphics[width=\columnwidth]{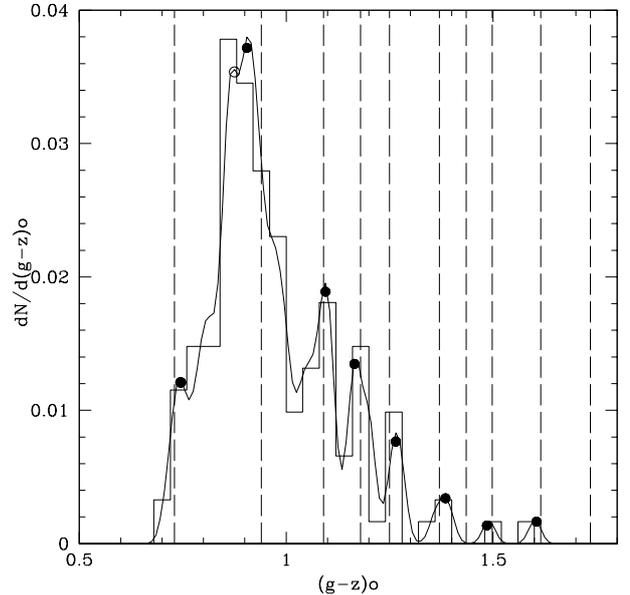}
    \caption{Smoothed (g-z)o colour distribution (adopting a 0.015 mag Gaussian kernel) corresponding to
 152 $GCs$ in six Fornax galaxies with $M_{g}$ -18.2 to -17.2 compared with the colours of the $TFP$
 (dashed lines; see text).
}
    \label{fig:fig19}
\end{figure}

 Fig.~\ref{fig:fig19} corresponds to 151 $GCs$ in six galaxies (FCC 55, 95, 143,
 152, 301, 335). Besides the prominent blue peak, there are other three well defined peaks at
 the expected position in the $TFP$ [3,4,5] with colour differences within $\pm$ 0.01 mag.
 We stress that these galaxies are not included in the sample
that defines the $TFP$ (which are about two magnitudes brighter). 
Remarkably, the four peaked pattern seen in this group survives when the
 sample is split in terms of apparent magnitude or galactocentric radius.
\subsubsection{GCCD for galaxies with $M_{g}=$-17.2 to -16.2}
\label{sss5b4}
\begin{figure}
	\includegraphics[width=\columnwidth]{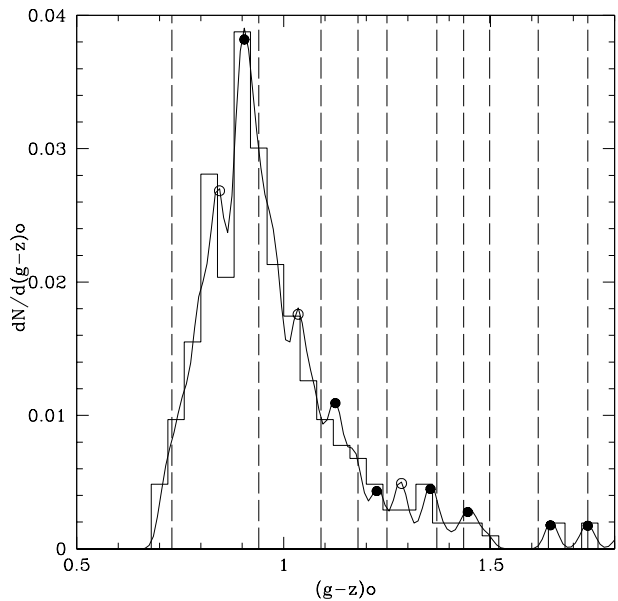}
    \caption{Smoothed (g-z)o colour distribution (adopting a 0.015 mag Gaussian kernel) corresponding to
 258 $GCs$ in 13 Fornax galaxies with $M_{g}$ -17.2 to -16.2 compared with the colours of the $TFP$ 
 (dashed lines; see text).
}
    \label{fig:fig20}
\end{figure}

 The $GCCD$ is shown in Fig.~\ref{fig:fig20} and includes 258 $GCs$ in 13 galaxies.
 This distribution is very similar to that corresponding to the previous galaxy
 group, except for the lack of any evident feature that could be compared with
 those in the $TVP$ (except the blue peak).
\section{The central galaxies in Virgo and Fornax}
\label{sec6}
In this section we discuss the results for $NGC~4486$
and $NGC~1399$ given their condition  of central  
galaxies in Virgo and Fornax, respectively, as well as
some preliminary results about other bright galaxies included
in the Appendix.

For these galaxies we performed an analysis similar
to that described in Section 4. In this case, the routine produces $GCCDs$
 within given ranges of galactocentric radii and position angles, and
looks for similarities with the corresponding template patterns.

A  similar analysis was performed on an independent photometric data
set of $GC$ candidates  both in $NGC~4486$ and $NGC~1399$ presented in
 \citet{Forte2007}.\\
Adopting a limiting magnitude $T_{1}$= 23.6 (that corresponds to $g$$\approx$ 24.5)
produces colour errors comparable to those of the $ACS$ photometry at $g$=25.0.
The  transformation of $(C-T_{1})$ to $(g-z)$ colours was derived from
a peripheral field in $NGC~4486$, that includes some 500 $GC$ candidates
with both Washington and $griz'$ photometry (\citealt{Forte2013}).
A revision of the data in this last paper leads to a bi-sector fit:
\begin{eqnarray}
\label{cero}
 (g-z) = 0.755 (C-T_{1}) - 0.058
\end{eqnarray}\\
(with slope and zero point errors of $\pm$ 0.073 and $\pm$ 0.012,
 respectively).\\
Taking into account the slope of this colour-colour relation,
we adopt a Gaussian kernel of 0.02 mag (instead of 0.015 mag) to obtain
 the smoothed $(C-T_{1})$ $GCCD$ and then transform this colour to
 $(g-z)$ through the last equation.
\subsection{Globular cluster colour distribution in NGC 4486.} 
\label{sbs6a}
\begin{figure}
	\includegraphics[width=\columnwidth]{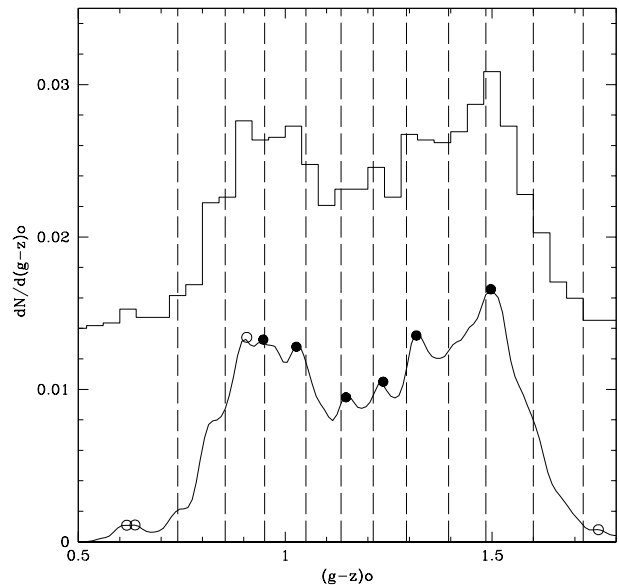}
    \caption{Discrete histogram (arbitrarily shifted upwards) and smoothed colour distribution for 1395 $GC$ candidates  with
 galactocentric radii from 0 to 120 $\arcsec$ in $NGC~4486$. Dashed lines correspond to the
$TVP$ (see text).
}
    \label{fig:fig21}
\end{figure}

Fig.~\ref{fig:fig21} displays the $GCCD$ for clusters in $NGC~4486$ both
in the discrete and smoothed ($\sigma_{(g-z)}=$ 0.015) histogram formats.
These $GCs$ (1395 objects) are brighter than $g=$25.0 and within an
annular region with galactocentric radii from 0 to 120 $\arcsec$. 
This figure shows a broad  bimodal structure with mild modulations 
that peak close to some of the colours that are defined in the $TVP$.

\begin{figure}
	\includegraphics[width=\columnwidth]{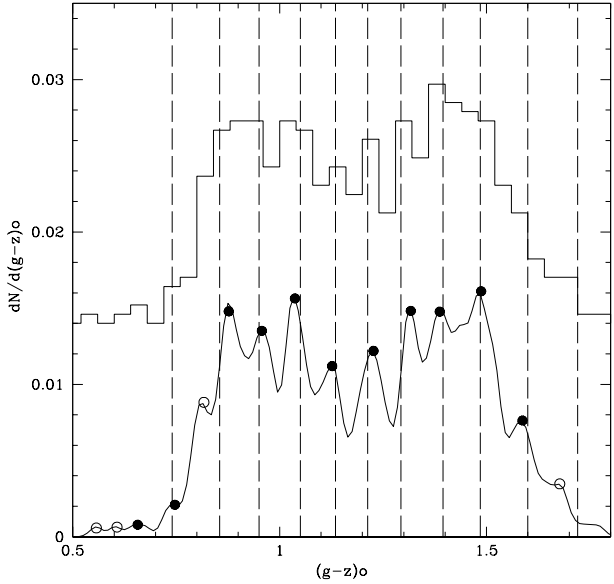}
    \caption{Discrete histogram (arbitrarily shifted upwards) and smoothed colour distribution for 414 $GC$ candidates in $NGC~4486$ 
 with galactocentric radii from 40 to 120 $\arcsec$ and position angles between 30 and 150 $\degr$.
 Dashed lines correspond to the $TVP$ (see text).
}
    \label{fig:fig22}
\end{figure}

Within this sample, the peak finding routine indicates that the best 
match to the $TVP$ occurs for $GCs$ in a galactocentric range from 40 to
120 $\arcsec$ and position angles (North to East) $PA$ between 30 and
150 $\degr$.
The $GCCD$ for 414 clusters in this region are shown in Fig.~\ref{fig:fig22}.
There are nine colour coincidences [2,3,4,5,6,7,8,9,10] with the
 $TVP$ to within $\pm$ 0.01 mag and without requiring any colour shift.
 
 The same analysis was carried out using the $(C-T_{1})o$ colours presented in 
 \citet{Forte2007}. These data, however, have a lower spatial coverage than that
 of the $ACS$ in the central region of the galaxy (i.e. galactocentric radii
 smaller than $\approx$ 40 $\arcsec$) due to image saturation.

\begin{figure}
	\includegraphics[width=\columnwidth]{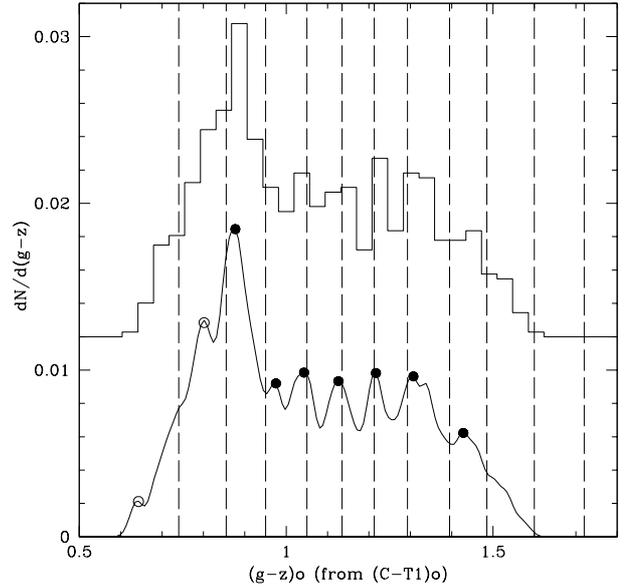}
    \caption{Discrete histogram (arbitrarily shifted upwards) and smoothed colour distribution for 683 $GCs$ in $NGC~4486$ with galactocentric radii
 between 90 and 150 $\arcsec$. Dashed lines correspond to the $TVP$ (see text).
}
    \label{fig:fig23}
\end{figure}

The $GCCD$ displayed in Fig.~\ref{fig:fig23}, corresponds to 683
$GCs$, with $(C-T_{1})o$ colours from 0.90 to 2.2, and within an annular region
 defined by galactocentric radii from 90 to 150 $\arcsec$. The best agreement with the
$TVP$ for peaks [2,3,4,5,6,7,8:] corresponds to a colour shift of -0.02 mag.

A further confirmation of the presence of the $TVP$ in $GCs$ associated
with $NGC~4486$ is provided by another independent data set. In this
case the $(g-z)'$ photometry comes from $Gemini-GMOS$ photometry given in
\citet{Forte2013} for clusters brighter than $g$ $\approx$ 23.4, and used to
 establish the  $(C-T_{1})$-$(g-z)$ relation presented in this section.

The $GCCD$ corresponding to this field (see Fig A5),
shows several peaks coincident with those in the $TVP$, [2,3,4:,5,6,7], with 
 colour differences within $\pm$ 0.01 mag.
It is worth mentioning that the presence of a multi-peak $GCCD$ in $NGC~4486$
  can be seen also in the $UV$ photometry presented by \citet{Bellini2015}.
 In particular, the $GCCD$ corresponding to their $(m_{F275W}-m_{F814W})$ colour
 index (see Fig. 6 in this last paper) also displays 6 to 8 peaks. Further
 analysis of this result, in the context of this paper, is currently under way.

Even though the results of other giant galaxies in Virgo, for which extensive
photometric data is available in the literature, will be discussed elsewhere
 (Forte et al. in prep.), some of them are included in the Appendix.
 For example, the $GCCD$ for the brightest galaxy in Virgo, $NGC~4472$
 exhibits nine colour peaks [2,3,4,5,6,8,9,10,11] within $\pm$ 0.015 mag from those
 defined by the $TVP$  (after applying a colour shift of -0.05 mag; see Fig A6).
 Peaks [1,7] are also detectable but exhibit colour differences larger than 0.02 mag.
 when compared with the nearest peak in the template pattern.
 This sample includes 517 $GCs$ with galactocentric radii from 20 to 110 $\arcsec$. 

\subsection{Globular cluster colour distribution in NGC 1399} 
\label{sbs6b}
\begin{figure}
	\includegraphics[width=\columnwidth]{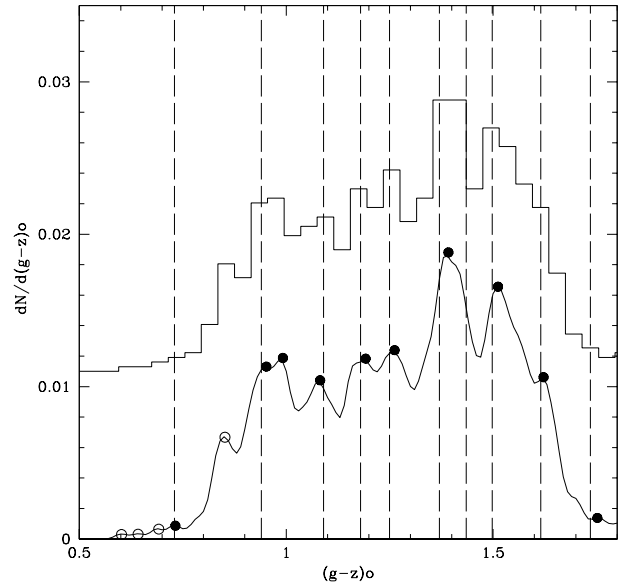}
    \caption{Discrete histogram (arbitrarily shifted upwards) and smoothed $(g-z)$ colour distribution for 814 $GCs$ in
 $NGC~1399$ with galactocentric radii from 0 to  120 $\arcsec$.
}
    \label{fig:fig24}
\end{figure}

The smoothed and discrete $GCCD$ for the giant galaxy $NGC~1399$, 
displayed in Fig.~\ref{fig:fig24}, corresponds to the $ACS$ photometry
of 814 $GCs$ within a galactocentric radius of 120 $\arcsec$,
 after adopting a colour shift of +0.01 mag.

\begin{figure}
	\includegraphics[width=\columnwidth]{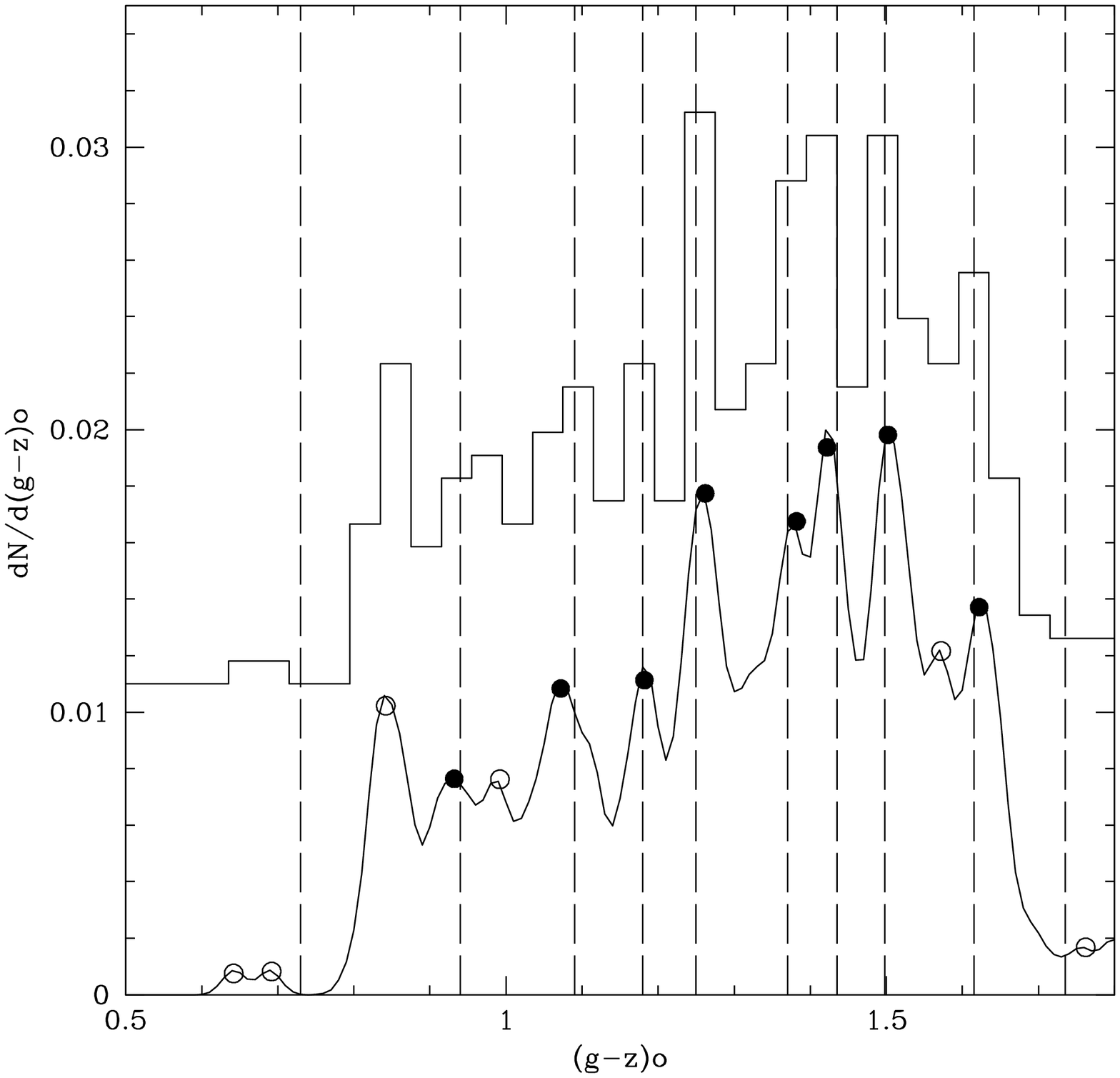}
    \caption{Discrete histogram (arbitrarily shifted upwards) and smoothed colour distribution for 309 $GCs$ in $NGC~1399$ with galactocentric
 radii from 0 to 90 $\arcsec$ and position angles between 90 and 270 $\degr$. Dashed lines correspond
 to the $TFP$ (see text).
}
    \label{fig:fig25}
\end{figure}

The coincidences with the $TFP$  are more evident for a                            
 sub-sample that includes 309 objects with galactocentric radii from 0
 to 90 $\arcsec$ and $PA$ from 90 to 270 $\degr$ (i.e. mainly to the south of
 the galaxy centre) shown in Fig.~\ref{fig:fig25}. The $GCCD$, after applying
 a colour shift of +0.02 mag., exhibits eight coincidences [2,3,4,5,6,7,8,9]
 to within $\pm$ 0.015 mag with the $TFP$. 

\begin{figure}
	\includegraphics[width=\columnwidth]{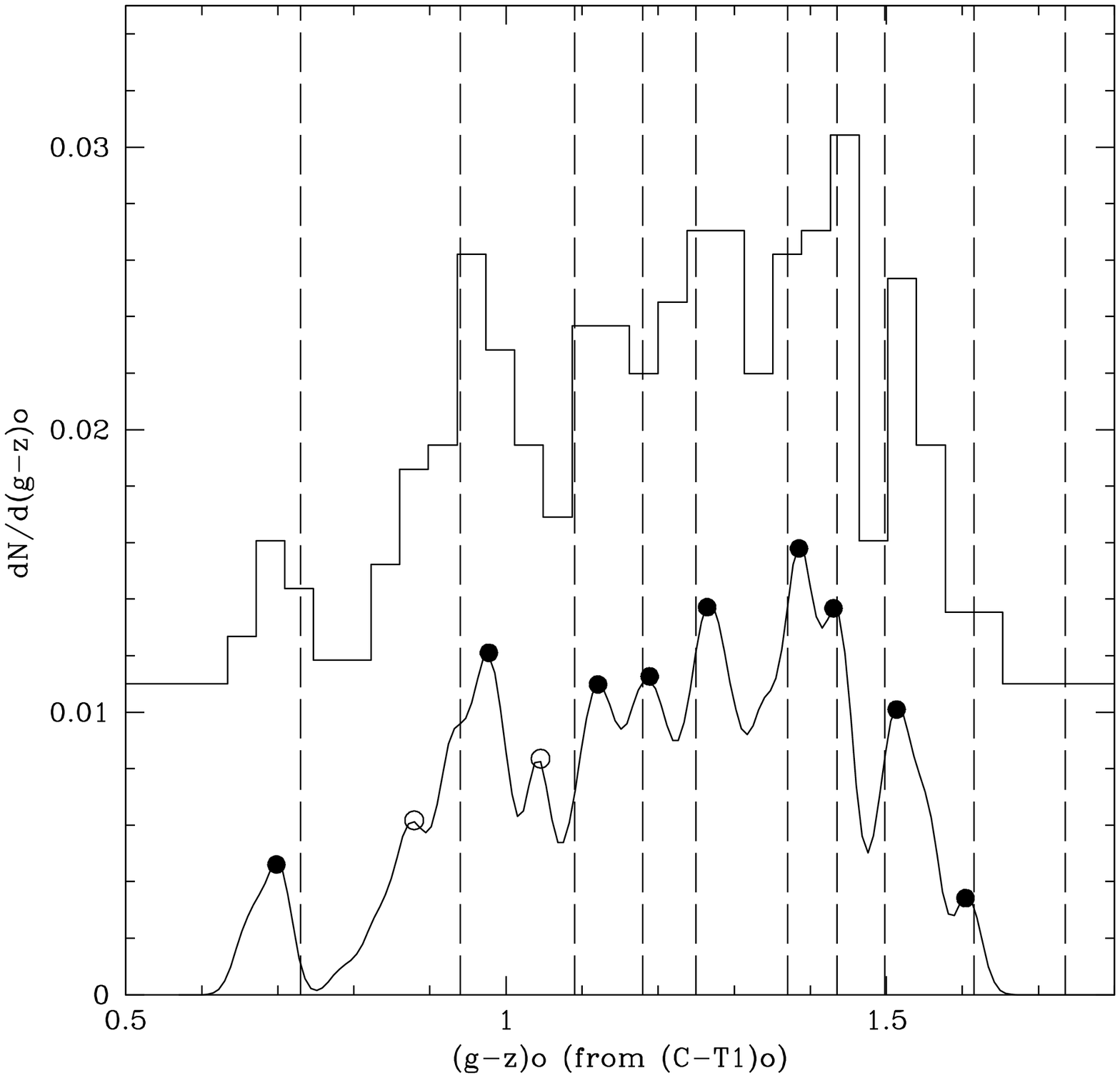}
    \caption{Discrete histogram (arbitrarily shifted upwards) and smoothed colour distribution for 293 $GCs$ in $NGC~1399$ with galactocentric
 radii from 0 to 120 $\arcsec$ and position angles between 90 and 360 $\degr$. Dashed lines correspond
 to the $TFP$ (see text).
}
    \label{fig:fig26}
\end{figure}

In turn, Fig.~\ref{fig:fig26} presents the $GCCD$ derived from $(C-T_{1})$ colours,
including 293 clusters in a galactocentric range from 0 to 120 $\arcsec$
and $PA$ 90 to 360 $\degr$. As in the previous diagram, several coincidences 
 with the $TFP$ can be detected to within $\pm$ 0.015: [3,4,5,6,7,8,9]. Peaks
 [1,2] are also detectable but exhibit colour differences larger than 0.025 mag.
 when compared with the nearest peak in the template pattern. This $GCs$ sample
 partially overlaps with that corresponding to the previous diagram but, due to
 image saturation, is incomplete below a galactocentric radius of 
 $\approx$ 50 $\arcsec$.

Other relevant cases, among the bright galaxies in Fornax, are
$NGC~1404$, located $\approx$ 15 $\arcmin$ from $NGC~1399$ on the sky, and $NGC~1316$, the
brightest galaxy in the Fornax cluster. This galaxy has a complex $GCS$ including several
cluster populations \citep{Goudfrooij2001a, Richtler2012b, Sesto2016}.
 
The $GCCD$ of the first of these galaxies exhibits nine coincidences [1,2,3,4,5,6,7,8,10]
 with the $TFP$ (see fig. $A 7$) that show up on top of what seems to be
 a rather bimodal colour distribution, a situation that is similar to that
 of the $GCCD$ in $NGC~4472$ (see Fig. A6).
Finally, Fig. A8  shows the $GCCD$ for a sample of 367 $GCs$ with galactocentric 
radii from 40 to 120 $\arcsec$ in $NGC~1316$.  Again, and top of a blue ward skewed 
colour distribution, there are several peaks coincident with the $TFP$. 


\section{Tentative multi-population fits}
\label{sec7}
Bimodal $GC$ colour distributions are usually represented in terms
of Gaussians that fit the blue and red $GC$ populations as shown, for
example, in \citet{Harris2017}.
A different approach links chemical abundance,
and $GC$ integrated colours assuming exponential dependences of the
number of clusters with $(Z/Z_{\odot})$ and colours 
 \citep[e.g.][]{VanDalfsen2004, Forte2007}.

 Alternatively, the appearance the composite $GCCDs$ for galaxies with
 $M_{g}=$ -20.2 to -19.2 suggests that these distributions could be matched
 in an $\it extreme~case$, just in terms of discrete mono-colour $GC$ populations.

In order to address this issue, we first determined the photometric
behaviour of a mono-colour $GC$ family, after adding the photometric
errors and the effect of the Gaussian smoothing kernel. In what
follows, we identify this function as the "colour spread function"
 ($CSF$).

Each colour corresponds, for a given age, to a chemical abundance
 $[Z/H]$. In this approach we adopt the colour-metallicity relation 
defined by the $SSP$ models presented by \citet{Bressan2012}, 
for an age of 12 Gy (but see Section 8), and a Salpeter integrated $GC$
 luminosity function.

These models, after a small correction to the $g$ magnitudes 
\citep[see][]{Sesto2016}, give a good match to the $GC$ multi-colour
 relations observed in the peripheral field of $NGC~4486$ (\citet{Forte2013}.

Photometric errors as a function of the $g$ and $z$ magnitudes were 
modelled adopting the values given in Table 4 and Table 2 of
\citet{Jordan2009} and \citet{Jordan2015} for $GCs$ in Virgo and
Fornax galaxies.

For galaxies with $M_{g}=$ -20.2 to -19.2 the $GC$ integrated luminosity
 function can be characterized by a Gaussian with a $\sigma$ parameter
of $\approx$ 1.0 mag. \citep{Villegas2010} and  turn-overs at 
 $g$=23.85 (Virgo) and $g=$24.26 (Fornax).

The procedure  starts generating Monte-Carlo apparent $g$ magnitudes,
controlled by the Gaussian integrated luminosity function, in a magnitude
range from $g=$ 20.0 to 25.0 and with a given metallicity. Once a model
$(g-z)$ colour is obtained from the colour-metallicity relation, a 
$z$ magnitude is derived. Model errors for  both the $g$ and $z$ magnitudes
are added, and an "observable" colour is derived from the magnitudes
difference. These  colours were, in turn, smoothed with a 0.015 mag 
Gaussian kernel (adopted in the analysis  described in previous 
Sections).

\begin{figure}
	\includegraphics[width=\columnwidth]{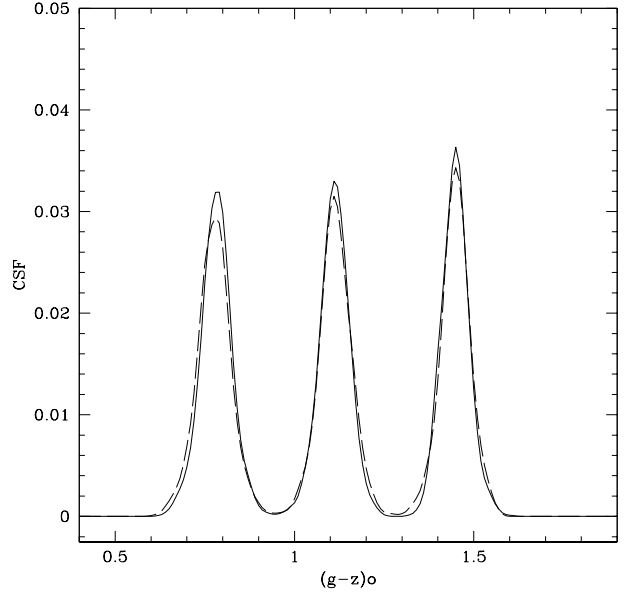}
    \caption{Colour spread function (CSF) for $GCs$ in Fornax (dashed lines) and Virgo (continuous lines) 
 corresponding to three chemical  abundances (left to right: [Z/H]=-2.0, -0.70 and 0.00). These 
 functions include the effect of photometric errors and are convolved with a Gaussian kernel of 0.015 mag
 in $(g-z)$.
}
    \label{fig:fig27}
\end{figure}

The resulting  $CSF$ functions for the Virgo and
Fornax datasets are depicted in Fig.~\ref{fig:fig27}.
 for three reference $[Z/H]$ values (-2.0; -0.7; 0.0).

The  fitting routine initially assigns equal amplitudes to all
components, and mean colours coincident with those of the
observed peaks in the corresponding $GCCD$. After computing the observed minus model
difference, each component is changed iteratively allowing for a variation
of its initial amplitude and colour. The adopted solution is 
the one  that minimizes the $rms$ of the observed minus model difference.
The adopted model, finally, is the mean of ten different realizations in
an attempt to minimize the statistical noise of the models.

\begin{figure}
	\includegraphics[width=\columnwidth]{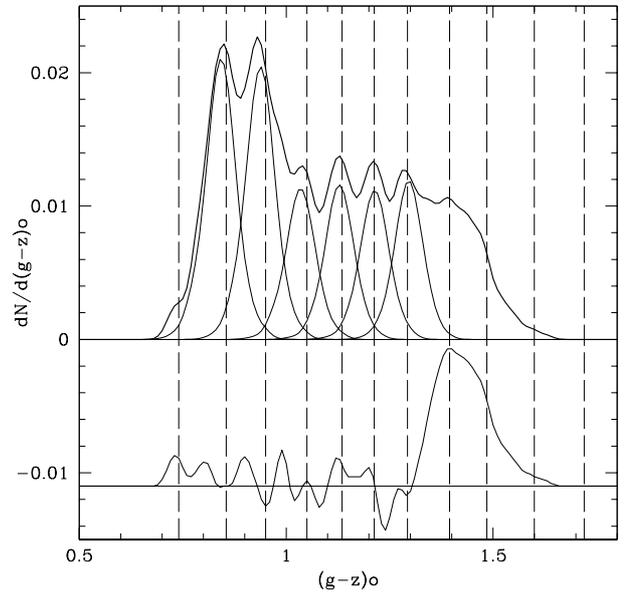}
    \caption{Smoothed colour distribution for $GCs$ defining the template pattern in Virgo galaxies
 (upper curve). The six components of the CSF are shown individually (thin lines). The observed
 minus model residuals are displayed in the lower curve (shifted by -0.011 in ordinates). Vertical
 lines indicate the $TVP$ colours. A broad red feature in the residual curve may be decomposed in  
 other two components present in the $TVP$. 
}
    \label{fig:fig28}
\end{figure}

The result for the Virgo galaxies is shown
in Fig.~\ref{fig:fig28}  and listed in Table~\ref{table_2}.
This case requires six discrete components. However, the fit suggests the presence of
 an extra component, at the reddest extreme of the distribution (shown in the residuals 
 curve), but not detectable as peaks in the composite $GCCD$ of these galaxies.
\begin{table}
\centering
\caption{Mono-colour 12 Gy $GC$ population fits in Virgo galaxies.}
\begin{tabular}{c c c c c}
\hline
\hline
\textbf{Number of GC}& &\textbf{[Z/H]}&\\
\hline
 270& &-1.65\\ 
 257& &-1.31\\
 142& &-0.93\\
 143& &-0.67\\
 137& &-0.46\\
 143& &-0.30\\
\hline
\label{table_2}
\end{tabular}
\end{table}
This red broad residual, in turn, overlaps with the position of two colour
peaks included in the $TVP$.

\begin{figure}
	\includegraphics[width=\columnwidth]{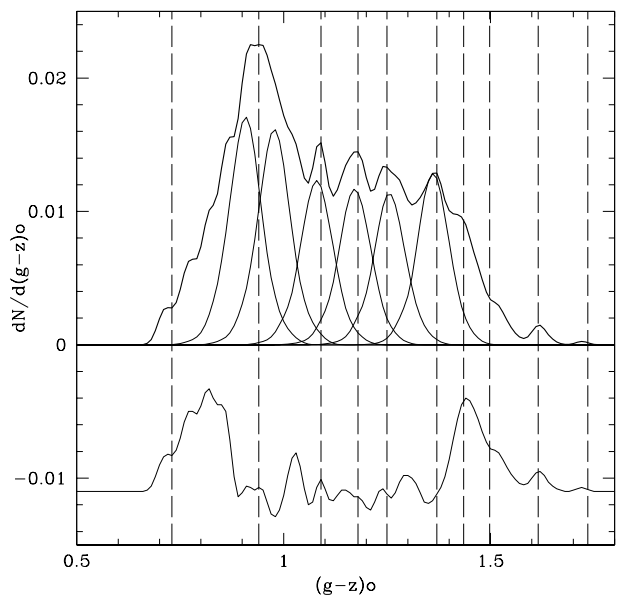}
    \caption{Smoothed colour distribution for $GCs$ defining the template pattern in Fornax galaxies (upper 
 curve). The six components of the CSF are shown individually (thin lines). The observed minus 
 model residuals are displayed in the lower curve (shifted by -0.011 in ordinates). Vertical lines 
 indicate the colours defined by the $TFP$ (see text). The two broad features in the blue
 and red extremes suggest the presence of other two components. 
}
    \label{fig:fig29}
\end{figure}

The  case of the $GCs$ associated with the counterpart galaxies in Fornax
is displayed in Fig.~\ref{fig:fig29} and listed in Table~\ref{table_3}.
For these galaxies the composite $GC$ colour
distribution and the $TFP$ shows a single broad blue peak. However, the
decomposition procedure requires at least two blue components.
The residuals curve shows that two more components (at the bluest and
reddest extremes) would be required for a proper fit
over the whole colour range. This indicates that, the broad and
single blue peak [2] detectable in the Fornax galaxies that define the $TFP$,
 may arise as the result of two to three ("non-resolved") mono-colour components.
\begin{table}
\centering
\caption{Mono-colour 12 Gy $GC$ population fits in Fornax galaxies.}
\begin{tabular}{c c c c c}
\hline
\hline
\textbf{Number of GC}& &\textbf{[Z/H]}&\\
\hline
 181& &-1.41\\ 
 171& &-1.15\\
 130& &-0.80\\
 123& &-0.56\\
 117& &-0.38\\
 132& &-0.17\\
\hline
\label{table_3}
\end{tabular}
\end{table}
The main conclusion of this Section is that, without discarding other
possible approaches, the  decomposition of the $GCCDs$  in terms of discrete
 mono-colours components cannot be rejected, at the level of the photometric
 errors, for galaxies with $M_{g}=$-20.2 to -19.2 both in Virgo and Fornax.
\section{Discussion}
\label{sec8}
The analysis presented in previous sections shows that colour
modulation patterns are present in the $GCCDs$ of composite $GC$
samples in some of the brightest Virgo and Fornax galaxies.

In principle, these modulations may have different origins.
Namely, field contamination, some kind of errors hidden in the photometric
data, statistical sampling noise or, finally, an alternative possibility 
that assumes that the template patterns have a physical entity.
 In what follows we comment on each of these items.

\subsection{Possible origin of the modulation pattern}
\subsubsection{Field contamination}
\label{sbs8b}
On one side, the discussion of this issue given in
\citet{Jordan2009} and \citet{Jordan2015}, indicates that
field contamination above the $GCs$ magnitude cut off 
adopted in this work ($g=$25.0) should be negligible.
On the other, the effect of eventual field contaminants should
be more important in the less populated composite $GCCDs$ of the fainter
galaxy groups. The results for these samples show that this
is not the case, i.e., field contamination can be safely ruled
out as tentative origin of the colour patterns.

\subsubsection{Photometric errors}
\label{sbs8c}
The appearance of the colour modulations might suggest the
presence of some kind of "periodic" errors (in colour terms) hidden in
 the $ACS$ photometric data. Such errors have not been reported in
the literature. On the other side, the template patterns are
 also detectable on independent data sets (e.g. the ground based Washington
 photometry and the $Gemini-GMOS$ photometry). Based on these
 results, the presence of eventual errors can be rejected. 
\subsubsection{Statistical noise}
\label{sbs8a}
The histograms corresponding to composite $GC$ samples in Virgo
and Fornax (for galaxies with $M_{g}=$-20.2 to -19.2) presented
initially in Fig.~ \ref{fig:fig9} and Fig.~ \ref{fig:fig10} show 
differences and similarities. For example, Virgo galaxies exhibit
a double blue peak [2,3] while their Fornax counterparts show
a single and broad blue peak [2]. In turn, both groups of galaxies 
present four to five colour peaks in the range occupied by
intermediate-to-red $GCs$.
 
The amplitude of these modulations is comparable to the
 expected Poissonian noise corresponding to the mean population 
 of intermediate-red $GCs$ per bin.
This argument might be considered as enough support in favour of
 statistical count fluctuations as the origin of the colour modulations.

Moreover, $GC$ populations modelling  based on a Monte Carlo approach
 \citep[e.g.][]{Forte2007}, can lead to statistical count fluctuations
 that, under given circumstances, show some similarities with the observed
 colour modulations. However, these models are $\it unable$ to replicate
 the survivability of the
 colour patterns when adopting different seed parameters or when
 the $GCCDs$ of the artificial clusters are analysed in subgroups
 characterized by ranges in magnitude (as shown in Fig. A2 and Fig. A4).

 An elementary statistical approach can be done assuming colour cells with
 sizes comparable to the $FWHM$ of the $CSFs$ ($\approx$0.1 mag.) discussed
 in the previous Section. The most evident (6 to 7) colour peaks in the $GCCDs$ fall 
 within a $(g-z)$ colour range of 0.8 mag (0.8 to 1.6) then comprising
 eight of these cells. With these parameters, the combinatorial number leads to 8
 different possible patterns (with 7 peaks),i.e., a particular one has a 12 percent probability
 of arising as the result of a random process. However, if the colour pattern can be 
 recognized after spliting the $GC$ sample, the random probability is decreased to only 1.5
 percent.  

 Furthermore, the $\it distinct$ template colour patterns emerged from the $\it composite$~$GCCDs$
 of galaxies within a given range of absolute magnitudes ($M_{g}=$-20.2 to -19.2)
 and were later found in the $\it individual$~$GCCDs$ in cluster sub-samples in
 the most massive elliptical galaxies: $NGC~4472$ and $NGC~4486$ in Virgo, and $NGC~1399$
 and $NGC~1404$ in Fornax. In fact the brightest galaxy in Fornax is $NGC~1316$ and,
 even in this complex system, some features of the $TFP$ are detectable (see Fig. A8).

 On the basis of all these arguments, stochastic effects can be rejected
 as the origin of the colour patterns. 
\subsubsection{An alternative possibility}
\label{sbs8d}     
 After rejecting  the eventual possibilities described above, we assume 
 that the colour patterns are real, distinct in Virgo and Fornax, and 
 suggest the existence of a previously unknown feature common to $GCS$
 in both galaxy clusters. A tentative explanation involves the effect of outer 
 stimuli leading to the enhancement of the $GCs$ formation process
 on different galaxies, in a synchronized way, as discussed below.
 
The main underlying questions are then, how such a features may
emerge on supra-galactic scales, and why they can be detected
more easily in galaxies with a given range in $M_{g}$ ?

Leaving the first question for the next sub-section, we can argue that
galaxies in the magnitude range from $M_{g}$ -20.2 to -19.2 seem 
somewhat "privileged" systems for the detection of the template
patterns: They have significant number of $GCs$ and, at the same time, 
have reached chemical enrichment levels as to include a considerable
number of red $GCs$.

The discussion presented in the previous Section in fact suggests that
 outer stimuli (enhancing the formation of mono-colour cluster
 populations) may have contributed to an important fraction of the
 total $GCS$ in these galaxies.

Fainter and less massive galaxies reach lower 
chemical  enrichment levels and would have not been able to show
evident features in the domain of the red $GCs$ (except a few
 cases noted previously (e.g. Fig.~\ref{fig:fig19}).

Invoking a "downsizing" scenario, most of these galaxies would be younger than
the brighter and more massive counterparts, and formed at redshifts 
 $z$$ \approx$ 2 or lower \citep[see][]{Paulino2017}, i.e., after the era
 when most of the energetic events took place (in the tentative time scale defined
 by $MW$-$GCs$).

In the case of $NGC~4486$ and $NGC~1399$, and other massive galaxies, their
rich $GCS$ presumably include clusters formed in a variety of processes that 
characterize their own histories. The colour patterns imposed by a putative 
external mechanism, would be overlapped in colour to those clusters but probably
exhibiting differences in age.  \citet{Usher2015} report that,
 in their analysis of the $GC$ colour-metallicity relations in different galaxies,
 these authors find a spread  possibly compatible with differences in $GC$ ages.

If the colour patterns are  the result of external events,
a chronological scale could be set on the basis that chemical abundance
$[Z/H]$ correlates with time. For example, \citet{Forbes2015} show that
blue and red $GCs$ exhibit substantial age differences, being the former
about 1.5 $Gy$ older.
In an $\it a~priori$ way, the $GCCD$ statistics may reflect either
a temporal sequence (where $[Z/H]$ is a "clock"), a positional sampling (i.e. halo, bulge),
 or a combination of both. This seems to be the case, for example, of $GCs$ in the $MW$.

 These clusters show a bifurcated $age$-$[Z/H]$ relation corresponding to halo
 or bulge-disc $GCs$ \citep[see][]{Leaman2013}. Both branches of that
 relation exhibit similar slopes in the $age$-$[Z/H]$ plane and an offset
 of 0.6 $dex$ in chemical abundance (in the sense that bulge-disc clusters have
 higher $[Z/H]$).

The existence of such relations in other galaxies is only an 
 arguable guess and some differences can be expected. For example,
more massive galaxies may display shallower slopes (eventually due to 
faster chemical enrichment) then compressing the time scale. This is
an important and obvious caveat in the following analysis.

\begin{figure}
	\includegraphics[width=\columnwidth]{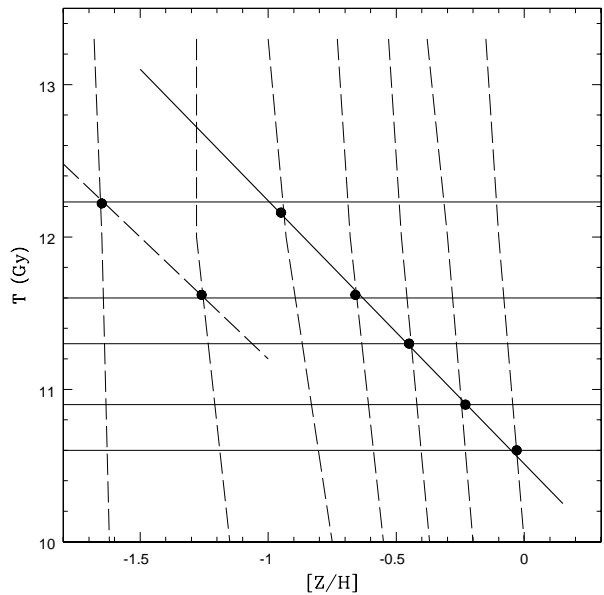}
    \caption{The bifurcated age-metallicity relation for halo (thick dashed line) and bulge-disk (thick solid
 line) $GCs$ in the $MW$. Thin dashed lines represent constant model  $(g-z)_{o}$ colours corresponding
 to  seven of the colour peaks observed in Virgo $GCs$ ([2,3,4,5,6,7,8]; see text). The interception of the age-metallicity
 relations with the  constant colour lines indicate the age of eventual external events triggering
 $GC$ formation (horizontal lines). The most recent events (filled dots) would impact only on 
 the bulge-disc $GCs$. The events span a range of $\approx$ 1.4 Gy.
}
    \label{fig:fig30}
\end{figure}

\begin{figure}
	\includegraphics[width=\columnwidth]{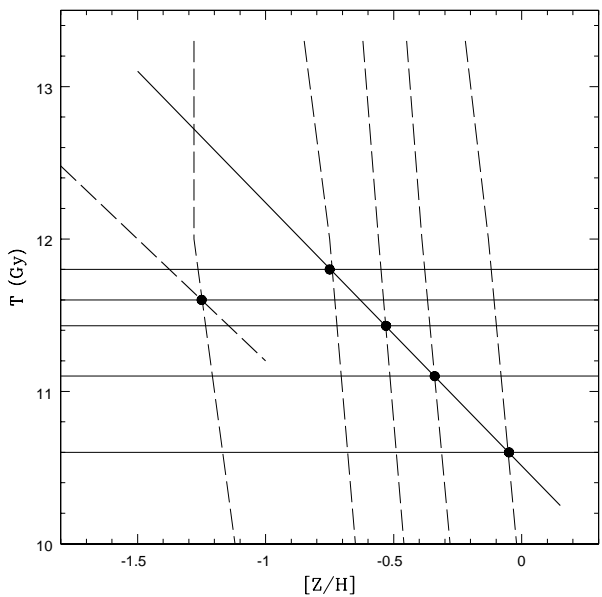}
    \caption{The bifurcated age-metallicity relation for halo (thick dashed line) and bulge-disk (thick solid
 line) $GCs$ in the $MW$. Thin dashed lines represent constant model $(g-z)_{o}$ colours corresponding
 to  five  of the colour peaks observed in Fornax $GCs$ ([2,3,4,5,6]; see text). The interception of the age-metallicity 
 relations with the  constant colour lines indicate the age of (eventual) external events triggering 
 $GC$ formation  (horizontal lines). The most recent events (filled dots) would impact only
 on the bulge-disc $GCs$. The events span a range in time of $\approx$ 1.2 Gy.
}
    \label{fig:fig31}
\end{figure}

The bifurcated $age$-$[Z/H]$ relations for $MW$ clusters are shown as thick 
dashed lines (blue halo $GCs$) and thick solid lines (bulge-disc $GCs$ in
Fig.~ \ref{fig:fig30} and Fig.~ \ref{fig:fig31}. Thin dashed lines in these
diagrams correspond to the most prominent features in the $TVP$ [2,3,4,5,6,7,8] 
and $TFP$ [2,3,4,5,6] patterns listed in Table~ \ref{table_2}.

The constant colour lines were derived from $SSP$ models by 
\citet{Bressan2012} and include a small correction to the $g$
magnitudes (that depend on metallicity) as described in \citet{Sesto2016}.

The interception between the $MW$~$age$-$[Z/H]$ relations and these lines
 would indicate the age (thin horizontal lines)  as well as the level of
 chemical enrichment of the interstellar medium in a galaxy at the time of a given 
 event leading to the enhancement of the $GC$ formation. The "width" of each of the
 colour features that define the template patterns is in fact dominated by 
 photometric errors (as dicussed in Section 7). This is a limiting factor for an
 estimate of the time lapse involved in each of these events. 

 Fig.~ \ref{fig:fig30} and Fig.~ \ref{fig:fig31} also show that the two first events
 would have had a simultaneous impact on both $GC$ metallicity sequences,
eventually leading to the formation of two peaks separated in colour.

The time lapse of these events range from $\approx$ 0.2 to 0.6 $Gy$.
A comparison between Fig.~\ref{fig:fig30} and Fig.~\ref{fig:fig31} indicates
that the earliest events in Virgo would have taken place some 0.4 Gy
earlier than in Fornax.

\subsection{Possible origin of the mono-colour GC populations}
\subsubsection{Super Massive Black Holes}
The role of Super Massive Black Holes ($SMBHs$) in this context remains
to be clarified. Interestingly, \citet*{HarrisG2014} 
find a strong correlation between the dynamical mass of galaxy bulges
and the number of associated clusters and stress that a causal $GCs-SMBH$ 
link cannot be simply dismissed. Both Virgo and Fornax have a number
 $SMBH$ (see Table 1 in the Harris et al. work) that might have been potential 
 players in generating high energy outflows producing an impact on
 the star forming process and, in particular, on the $GC$ formation.
\subsubsection{Cluster-Cluster mergers}
A number of other possible mechanisms deserve to be explored. 
For example, in  $\Lambda$-CDM cosmologies, galaxy clusters
are the result of the merger of smaller sub-clusters. An outstanding
 object in this frame is $CIZAJ22428+5301$ \citep{Sobral2015, Stroe2015}, an
old galaxy cluster at $z=$ 0.19, with a violent stellar formation activity fuelled
by a galaxy cluster merger. This activity takes place over a diameter of 
 $\approx$ 3 $Mpc$ a spatial scale that is considerably larger than, for example,
 the core radii of the Virgo and Fornax clusters (in the order of $\approx$ 1 $Mpc$).

A possible scenario suggests that, the strong shock waves originated in 
sub-cluster mergers at high red-shifts, may have favoured the formation of
 $GCs$ if the  proper sites (dark matter mini-haloes ?) are available in a
galaxy. The eventual availability of these places might explain the lack
of a clear regularity in the amplitude of the colour patterns in different
$GCS$.\\
\section{Conclusions}
\label{sec9}
The nature of the distinct colour modulations in the $GCCDs$ of clusters
associated  with Virgo and Fornax galaxies, and characterized through
the so called "template patterns" ($TVP$ and $TFP$), is an intriguing question.
These structures are no evident in currently available spectroscopic data,
possibly as a result of the still relatively large errors in age and chemical
abundance.

At this point, the main argument to support their $\it entity$ as features 
with a physical meaning, stands on the fact that, once recognized in a given sub-group
of galaxies that defined the template colour patterns, they can be
 found also in the rich $GCS$ of the central galaxies  $NGC~4486$ and
 $NGC~1399$, as well as in other bright galaxies (not included in the
 galaxy groups that define the template patterns).

The $GCS$ associated with these massive galaxies reflect their own
and complex histories including, for example, the accretion of $GCs$ belonging
to less massive systems during mergers \citep*{Forte1982, Cote1998}. 
 We note that, in these processes, dry mergers involving galaxies that already have
 incorporated the template pattern structures, would preserve these features.

The physical origin of these colour patterns, shared by a number of galaxies
spread on a large spatial volume, at this stage is clearly speculative. However,
their detection has been the result of the $\it motivation$ of this work given in
the Introduction: The search for detectable signatures, possibly connected with 
energetic events, on the $GCs$ formation process.
 
An important and open issue connected with $GCs$ is why they do not form 
currently (except, possibly, in on-going galaxy mergers) despite that
 enough interstellar matter is available to feed such
a process (for example, in the $MW$). The missing ingredient can be
(cautiously) identified with those events.

The spatial positions of the galaxies that define the
template colour patterns do not show an evident systematic distribution, i.e.,
these galaxies are spread over the whole  Virgo and Fornax clusters.
The  positions of these galaxies, however, do not necessarily
reflect the conditions prevailing at the high red-shift regime when
the energetic events eventually took place.

In any case, it is clear that
these phenomena should have been powerful enough to leave an 
imprint on large spatial volumes and possibly not only on $GCS$ but
also on the galaxy stellar populations in general. For example,
ultra compact dwarfs (UCDs) include distinct stellar sub populations with
rather defined ages and chemical abundances \citep[e.g.][]{Norris2015} that
may also fit in that landscape.

Rather than stellar formation "quenching" the (peaked) colour patterns
suggest that the putative phenomenon has been able to enhance the formation
 of $GCs$ . Several mechanisms that may increase the star formation rate have
 been proposed in the recent literature \citep*{NZ2012, Silk2013, Abe2016}

 The similarities of the colour patterns in Virgo and Fornax might be
 related to environmental conditions (i.e., these are galaxy clusters with
 dominant giant galaxies).
 
 The detectability of these features in other environments, e.g., the Local
 Group, with much smaller $GC$ populations will be presumably hard to asses. However,
 it is possible that large scale events may leave an imprint not only
 on the $GCCDs$ but also on individual $GCs$.

 For example, the presence of multi populations in $MW$ clusters, as the remarkable case
 of $Terzan~5$ \citep[see][]{Ferraro2016}, may provide some clues about the eventual impact of 
 those phenomena on the life of galaxies located in poorly populated environments.

 Finally, as an unavoidable conclusion, the analysis of the $GCs$ colour
 patterns will require a considerable volume of high precision photometry
and spectroscopy (involving very low errors both in age and $[Z/H]$) for a complete
 characterization of their astrophysical properties.
 
\section*{Acknowledgements}
This work was funded with grants from Consejo Nacional de Investigaciones
Cientificas y Tecnicas de la Republica Argentina $(CONICET)$. Thanks are also due
 to Dr. E. Irene Vega for her continuous support during the redaction of this work. \\

\bibliographystyle{mnras}
\bibliography{biblio_Forte.bib} 

\begin{thebibliography}{}
\makeatletter
\relax
\def\mn@urlcharsother{\let\do\@makeother \do\$\do\&\do\#\do\^\do\_\do\%\do\~}
\def\mn@doi{\begingroup\mn@urlcharsother \@ifnextchar [ {\mn@doi@}
  {\mn@doi@[]}}
\def\mn@doi@[#1]#2{\def\@tempa{#1}\ifx\@tempa\@empty \href
  {http://dx.doi.org/#2} {doi:#2}\else \href {http://dx.doi.org/#2} {#1}\fi
  \endgroup}
\def\mn@eprint#1#2{\mn@eprint@#1:#2::\@nil}
\def\mn@eprint@arXiv#1{\href {http://arxiv.org/abs/#1} {{\tt arXiv:#1}}}
\def\mn@eprint@dblp#1{\href {http://dblp.uni-trier.de/rec/bibtex/#1.xml}
  {dblp:#1}}
\def\mn@eprint@#1:#2:#3:#4\@nil{\def\@tempa {#1}\def\@tempb {#2}\def\@tempc
  {#3}\ifx \@tempc \@empty \let \@tempc \@tempb \let \@tempb \@tempa \fi \ifx
  \@tempb \@empty \def\@tempb {arXiv}\fi \@ifundefined
  {mn@eprint@\@tempb}{\@tempb:\@tempc}{\expandafter \expandafter \csname
  mn@eprint@\@tempb\endcsname \expandafter{\@tempc}}}

\bibitem[\protect\citeauthoryear{{Abe}, {Umemura}  \& {Hasegawa}}{{Abe}
  et~al.}{2016}]{Abe2016}
{Abe} M.,  {Umemura} M.,   {Hasegawa} K.,  2016, \mn@doi [\mnras]
  {10.1093/mnras/stw2164}, \href
  {http://adsabs.harvard.edu/abs/2016MNRAS.463.2849A} {463, 2849}

\bibitem[\protect\citeauthoryear{{Bellini} et~al.,}{{Bellini}
  et~al.}{2015}]{Bellini2015}
{Bellini} A.,  et~al., 2015, \mn@doi [\apj] {10.1088/0004-637X/805/2/178},
  \href {http://adsabs.harvard.edu/abs/2015ApJ...805..178B} {805, 178}

\bibitem[\protect\citeauthoryear{{Binggeli}, {Sandage}  \&
  {Tarenghi}}{{Binggeli} et~al.}{1984}]{Binggeli1984}
{Binggeli} B.,  {Sandage} A.,   {Tarenghi} M.,  1984, \mn@doi [\aj]
  {10.1086/113484}, \href {http://adsabs.harvard.edu/abs/1984AJ.....89...64B}
  {89, 64}

\bibitem[\protect\citeauthoryear{{Blakeslee} et~al.,}{{Blakeslee}
  et~al.}{2009}]{Blakeslee2009}
{Blakeslee} J.~P.,  et~al., 2009, \mn@doi [\apj] {10.1088/0004-637X/694/1/556},
  \href {http://adsabs.harvard.edu/abs/2009ApJ...694..556B} {694, 556}

\bibitem[\protect\citeauthoryear{{Bressan}, {Marigo}, {Girardi}, {Salasnich},
  {Dal Cero}, {Rubele}  \& {Nanni}}{{Bressan} et~al.}{2012}]{Bressan2012}
{Bressan} A.,  {Marigo} P.,  {Girardi} L.,  {Salasnich} B.,  {Dal Cero} C.,
  {Rubele} S.,   {Nanni} A.,  2012, \mn@doi [\mnras]
  {10.1111/j.1365-2966.2012.21948.x}, \href
  {http://adsabs.harvard.edu/abs/2012MNRAS.427..127B} {427, 127}

\bibitem[\protect\citeauthoryear{{Brodie} \& {Strader}}{{Brodie} \&
  {Strader}}{2006}]{Brodie2006}
{Brodie} J.~P.,  {Strader} J.,  2006, \mn@doi [\araa]
  {10.1146/annurev.astro.44.051905.092441}, \href
  {http://adsabs.harvard.edu/abs/2006ARA%26A..44..193B} {44, 193}

\bibitem[\protect\citeauthoryear{{Cen}}{{Cen}}{2001}]{Cen2001}
{Cen} R.,  2001, \mn@doi [\apj] {10.1086/323071}, \href
  {http://adsabs.harvard.edu/abs/2001ApJ...560..592C} {560, 592}

\bibitem[\protect\citeauthoryear{{Chen}, {C{\^o}t{\'e}}, {West}, {Peng}  \&
  {Ferrarese}}{{Chen} et~al.}{2010}]{Chen2010}
{Chen} C.-W.,  {C{\^o}t{\'e}} P.,  {West} A.~A.,  {Peng} E.~W.,   {Ferrarese}
  L.,  2010, \mn@doi [\apjs] {10.1088/0067-0049/191/1/1}, \href
  {http://adsabs.harvard.edu/abs/2010ApJS..191....1C} {191, 1}

\bibitem[\protect\citeauthoryear{{Chonis} \& {Gaskell}}{{Chonis} \&
  {Gaskell}}{2008}]{Chonis2008}
{Chonis} T.~S.,  {Gaskell} C.~M.,  2008, \mn@doi [\aj]
  {10.1088/0004-6256/135/1/264}, \href
  {http://adsabs.harvard.edu/abs/2008AJ....135..264C} {135, 264}

\bibitem[\protect\citeauthoryear{{C{\^o}t{\'e}}, {Marzke}  \&
  {West}}{{C{\^o}t{\'e}} et~al.}{1998}]{Cote1998}
{C{\^o}t{\'e}} P.,  {Marzke} R.~O.,   {West} M.~J.,  1998, \mn@doi [\apj]
  {10.1086/305838}, \href {http://adsabs.harvard.edu/abs/1998ApJ...501..554C}
  {501, 554}

\bibitem[\protect\citeauthoryear{{D'Abrusco}, {Fabbiano}  \&
  {Zezas}}{{D'Abrusco} et~al.}{2015}]{Dabrusco2015}
{D'Abrusco} R.,  {Fabbiano} G.,   {Zezas} A.,  2015, \mn@doi [\apj]
  {10.1088/0004-637X/805/1/26}, \href
  {http://adsabs.harvard.edu/abs/2015ApJ...805...26D} {805, 26}

\bibitem[\protect\citeauthoryear{{Fabian}}{{Fabian}}{2012}]{Fabian2012}
{Fabian} A.~C.,  2012, \mn@doi [\araa] {10.1146/annurev-astro-081811-125521},
  \href {http://adsabs.harvard.edu/abs/2012ARA%26A..50..455F} {50, 455}

\bibitem[\protect\citeauthoryear{{Faifer} et~al.,}{{Faifer}
  et~al.}{2011}]{Faifer2011}
{Faifer} F.~R.,  et~al., 2011, \mn@doi [\mnras]
  {10.1111/j.1365-2966.2011.19018.x}, \href
  {http://adsabs.harvard.edu/abs/2011MNRAS.416..155F} {416, 155}

\bibitem[\protect\citeauthoryear{{Ferguson}}{{Ferguson}}{1989}]{Ferguson1989}
{Ferguson} H.~C.,  1989, \mn@doi [\aj] {10.1086/115152}, \href
  {http://adsabs.harvard.edu/abs/1989AJ.....98..367F} {98, 367}

\bibitem[\protect\citeauthoryear{{Ferraro}, {Massari}, {Dalessandro},
  {Lanzoni}, {Origlia}, {Rich}  \& {Mucciarelli}}{{Ferraro}
  et~al.}{2016}]{Ferraro2016}
{Ferraro} F.~R.,  {Massari} D.,  {Dalessandro} E.,  {Lanzoni} B.,  {Origlia}
  L.,  {Rich} R.~M.,   {Mucciarelli} A.,  2016, \mn@doi [\apj]
  {10.3847/0004-637X/828/2/75}, \href
  {http://adsabs.harvard.edu/abs/2016ApJ...828...75F} {828, 75}

\bibitem[\protect\citeauthoryear{{Forbes}, {Ponman}  \& {O'Sullivan}}{{Forbes}
  et~al.}{2012}]{Forbes2012}
{Forbes} D.~A.,  {Ponman} T.,   {O'Sullivan} E.,  2012, \mn@doi [\mnras]
  {10.1111/j.1365-2966.2012.21368.x}, \href
  {http://adsabs.harvard.edu/abs/2012MNRAS.425...66F} {425, 66}

\bibitem[\protect\citeauthoryear{{Forbes}, {Pastorello}, {Romanowsky}, {Usher},
  {Brodie}  \& {Strader}}{{Forbes} et~al.}{2015}]{Forbes2015}
{Forbes} D.~A.,  {Pastorello} N.,  {Romanowsky} A.~J.,  {Usher} C.,  {Brodie}
  J.~P.,   {Strader} J.,  2015, \mn@doi [\mnras] {10.1093/mnras/stv1312}, \href
  {http://adsabs.harvard.edu/abs/2015MNRAS.452.1045F} {452, 1045}

\bibitem[\protect\citeauthoryear{{Forbes}, {Alabi}, {Romanowsky}, {Brodie},
  {Strader}, {Usher}  \& {Pota}}{{Forbes} et~al.}{2016}]{Forbes2016}
{Forbes} D.~A.,  {Alabi} A.,  {Romanowsky} A.~J.,  {Brodie} J.~P.,  {Strader}
  J.,  {Usher} C.,   {Pota} V.,  2016, \mn@doi [\mnras]
  {10.1093/mnrasl/slw015}, \href
  {http://adsabs.harvard.edu/abs/2016MNRAS.458L..44F} {458, L44}

\bibitem[\protect\citeauthoryear{{Forte}, {Martinez}  \& {Muzzio}}{{Forte}
  et~al.}{1982}]{Forte1982}
{Forte} J.~C.,  {Martinez} R.~E.,   {Muzzio} J.~C.,  1982, \mn@doi [\aj]
  {10.1086/113236}, \href {http://adsabs.harvard.edu/abs/1982AJ.....87.1465F}
  {87, 1465}

\bibitem[\protect\citeauthoryear{{Forte}, {Faifer}  \& {Geisler}}{{Forte}
  et~al.}{2007}]{Forte2007}
{Forte} J.~C.,  {Faifer} F.,   {Geisler} D.,  2007, \mn@doi [\mnras]
  {10.1111/j.1365-2966.2007.12515.x}, \href
  {http://adsabs.harvard.edu/abs/2007MNRAS.382.1947F} {382, 1947}

\bibitem[\protect\citeauthoryear{{Forte}, {Vega}  \& {Faifer}}{{Forte}
  et~al.}{2012}]{Forte2012}
{Forte} J.~C.,  {Vega} E.~I.,   {Faifer} F.,  2012, \mn@doi [\mnras]
  {10.1111/j.1365-2966.2011.20341.x}, \href
  {http://adsabs.harvard.edu/abs/2012MNRAS.421..635F} {421, 635}

\bibitem[\protect\citeauthoryear{{Forte}, {Faifer}, {Vega}, {Bassino}, {Smith
  Castelli}, {Cellone}  \& {Geisler}}{{Forte} et~al.}{2013}]{Forte2013}
{Forte} J.~C.,  {Faifer} F.~R.,  {Vega} E.~I.,  {Bassino} L.~P.,  {Smith
  Castelli} A.~V.,  {Cellone} S.~A.,   {Geisler} D.,  2013, \mn@doi [\mnras]
  {10.1093/mnras/stt263}, \href
  {http://adsabs.harvard.edu/abs/2013MNRAS.431.1405F} {431, 1405}

\bibitem[\protect\citeauthoryear{{Forte}, {Vega}, {Faifer}, {Smith Castelli},
  {Escudero}, {Gonz{\'a}lez}  \& {Sesto}}{{Forte} et~al.}{2014}]{Forte2014}
{Forte} J.~C.,  {Vega} E.~I.,  {Faifer} F.~R.,  {Smith Castelli} A.~V.,
  {Escudero} C.,  {Gonz{\'a}lez} N.~M.,   {Sesto} L.,  2014, \mn@doi [\mnras]
  {10.1093/mnras/stu658}, \href
  {http://adsabs.harvard.edu/abs/2014MNRAS.441.1391F} {441, 1391}

\bibitem[\protect\citeauthoryear{{Goudfrooij}, {Mack}, {Kissler-Patig},
  {Meylan}  \& {Minniti}}{{Goudfrooij} et~al.}{2001}]{Goudfrooij2001a}
{Goudfrooij} P.,  {Mack} J.,  {Kissler-Patig} M.,  {Meylan} G.,   {Minniti} D.,
   2001, \mn@doi [\mnras] {10.1046/j.1365-8711.2001.04154.x}, \href
  {http://adsabs.harvard.edu/abs/2001MNRAS.322..643G} {322, 643}

\bibitem[\protect\citeauthoryear{{Harris}, {Poole}  \& {Harris}}{{Harris}
  et~al.}{2014}]{HarrisG2014}
{Harris} G.~L.~H.,  {Poole} G.~B.,   {Harris} W.~E.,  2014, \mn@doi [\mnras]
  {10.1093/mnras/stt2337}, \href
  {http://adsabs.harvard.edu/abs/2014MNRAS.438.2117H} {438, 2117}

\bibitem[\protect\citeauthoryear{{Harris}, {Ciccone}, {Eadie}, {Gnedin},
  {Geisler}, {Rothberg}  \& {Bailin}}{{Harris} et~al.}{2017}]{Harris2017}
{Harris} W.~E.,  {Ciccone} S.~M.,  {Eadie} G.~M.,  {Gnedin} O.~Y.,  {Geisler}
  D.,  {Rothberg} B.,   {Bailin} J.,  2017, \mn@doi [\apj]
  {10.3847/1538-4357/835/1/101}, \href
  {http://adsabs.harvard.edu/abs/2017ApJ...835..101H} {835, 101}

\bibitem[\protect\citeauthoryear{{Jord{\'a}n} et~al.,}{{Jord{\'a}n}
  et~al.}{2009}]{Jordan2009}
{Jord{\'a}n} A.,  et~al., 2009, \mn@doi [\apjs] {10.1088/0067-0049/180/1/54},
  \href {http://adsabs.harvard.edu/abs/2009ApJS..180...54J} {180, 54}

\bibitem[\protect\citeauthoryear{{Jord{\'a}n}, {Peng}, {Blakeslee},
  {C{\^o}t{\'e}}, {Eyheramendy}  \& {Ferrarese}}{{Jord{\'a}n}
  et~al.}{2015}]{Jordan2015}
{Jord{\'a}n} A.,  {Peng} E.~W.,  {Blakeslee} J.~P.,  {C{\^o}t{\'e}} P.,
  {Eyheramendy} S.,   {Ferrarese} L.,  2015, \mn@doi [\apjs]
  {10.1088/0067-0049/221/1/13}, \href
  {http://adsabs.harvard.edu/abs/2015ApJS..221...13J} {221, 13}

\bibitem[\protect\citeauthoryear{{Leaman}, {VandenBerg}  \& {Mendel}}{{Leaman}
  et~al.}{2013}]{Leaman2013}
{Leaman} R.,  {VandenBerg} D.~A.,   {Mendel} J.~T.,  2013, \mn@doi [\mnras]
  {10.1093/mnras/stt1540}, \href
  {http://adsabs.harvard.edu/abs/2013MNRAS.436..122L} {436, 122}

\bibitem[\protect\citeauthoryear{{Milone} et~al.,}{{Milone}
  et~al.}{2017}]{Milone2017}
{Milone} A.~P.,  et~al., 2017, \mn@doi [\mnras] {10.1093/mnras/stw2531}, \href
  {http://adsabs.harvard.edu/abs/2017MNRAS.464.3636M} {464, 3636}

\bibitem[\protect\citeauthoryear{{Nayakshin} \& {Zubovas}}{{Nayakshin} \&
  {Zubovas}}{2012}]{NZ2012}
{Nayakshin} S.,  {Zubovas} K.,  2012, \mn@doi [\mnras]
  {10.1111/j.1365-2966.2012.21950.x}, \href
  {http://adsabs.harvard.edu/abs/2012MNRAS.427..372N} {427, 372}

\bibitem[\protect\citeauthoryear{{Norris}, {Escudero}, {Faifer}, {Kannappan},
  {Forte}  \& {van den Bosch}}{{Norris} et~al.}{2015}]{Norris2015}
{Norris} M.~A.,  {Escudero} C.~G.,  {Faifer} F.~R.,  {Kannappan} S.~J.,
  {Forte} J.~C.,   {van den Bosch} R.~C.~E.,  2015, \mn@doi [\mnras]
  {10.1093/mnras/stv1221}, \href
  {http://adsabs.harvard.edu/abs/2015MNRAS.451.3615N} {451, 3615}

\bibitem[\protect\citeauthoryear{{Paulino-Afonso}, {Sobral}, {Buitrago}  \&
  {Afonso}}{{Paulino-Afonso} et~al.}{2017}]{Paulino2017}
{Paulino-Afonso} A.,  {Sobral} D.,  {Buitrago} F.,   {Afonso} J.,  2017,
  \mn@doi [\mnras] {10.1093/mnras/stw2933}, \href
  {http://adsabs.harvard.edu/abs/2017MNRAS.465.2717P} {465, 2717}

\bibitem[\protect\citeauthoryear{{Peng} et~al.,}{{Peng}
  et~al.}{2006}]{Peng2006}
{Peng} E.~W.,  et~al., 2006, \mn@doi [\apj] {10.1086/499485}, \href
  {http://adsabs.harvard.edu/abs/2006ApJ...639..838P} {639, 838}

\bibitem[\protect\citeauthoryear{{Richtler}, {Bassino}, {Dirsch}  \&
  {Kumar}}{{Richtler} et~al.}{2012}]{Richtler2012b}
{Richtler} T.,  {Bassino} L.~P.,  {Dirsch} B.,   {Kumar} B.,  2012, \mn@doi
  [\aap] {10.1051/0004-6361/201118589}, \href
  {http://adsabs.harvard.edu/abs/2012A%26A...543A.131R} {543, A131}

\bibitem[\protect\citeauthoryear{{Schlafly} \& {Finkbeiner}}{{Schlafly} \&
  {Finkbeiner}}{2011}]{Schlafly2011}
{Schlafly} E.~F.,  {Finkbeiner} D.~P.,  2011, \mn@doi [\apj]
  {10.1088/0004-637X/737/2/103}, \href
  {http://adsabs.harvard.edu/abs/2011ApJ...737..103S} {737, 103}

\bibitem[\protect\citeauthoryear{{Sesto}, {Faifer}  \& {Forte}}{{Sesto}
  et~al.}{2016}]{Sesto2016}
{Sesto} L.~A.,  {Faifer} F.~R.,   {Forte} J.~C.,  2016, \mn@doi [\mnras]
  {10.1093/mnras/stw1627}, \href
  {http://adsabs.harvard.edu/abs/2016MNRAS.461.4260S} {461, 4260}

\bibitem[\protect\citeauthoryear{{Silk}}{{Silk}}{2013}]{Silk2013}
{Silk} J.,  2013, \mn@doi [\apj] {10.1088/0004-637X/772/2/112}, \href
  {http://adsabs.harvard.edu/abs/2013ApJ...772..112S} {772, 112}

\bibitem[\protect\citeauthoryear{{Sobral}, {Stroe}, {Dawson}, {Wittman}, {Jee},
  {R{\"o}ttgering}, {van Weeren}  \& {Br{\"u}ggen}}{{Sobral}
  et~al.}{2015}]{Sobral2015}
{Sobral} D.,  {Stroe} A.,  {Dawson} W.~A.,  {Wittman} D.,  {Jee} M.~J.,
  {R{\"o}ttgering} H.,  {van Weeren} R.~J.,   {Br{\"u}ggen} M.,  2015, \mn@doi
  [\mnras] {10.1093/mnras/stv521}, \href
  {http://adsabs.harvard.edu/abs/2015MNRAS.450..630S} {450, 630}

\bibitem[\protect\citeauthoryear{{Stroe} et~al.,}{{Stroe}
  et~al.}{2015}]{Stroe2015}
{Stroe} A.,  et~al., 2015, \mn@doi [\mnras] {10.1093/mnras/stu2519}, \href
  {http://adsabs.harvard.edu/abs/2015MNRAS.450..646S} {450, 646}

\bibitem[\protect\citeauthoryear{{Usher} et~al.,}{{Usher}
  et~al.}{2015}]{Usher2015}
{Usher} C.,  et~al., 2015, \mn@doi [\mnras] {10.1093/mnras/stu2050}, \href
  {http://adsabs.harvard.edu/abs/2015MNRAS.446..369U} {446, 369}

\bibitem[\protect\citeauthoryear{{VanDalfsen} \& {Harris}}{{VanDalfsen} \&
  {Harris}}{2004}]{VanDalfsen2004}
{VanDalfsen} M.~L.,  {Harris} W.~E.,  2004, \mn@doi [\aj] {10.1086/380227},
  \href {http://adsabs.harvard.edu/abs/2004AJ....127..368V} {127, 368}

\bibitem[\protect\citeauthoryear{{Vanzella} et~al.,}{{Vanzella}
  et~al.}{2016}]{Vanzella2016}
{Vanzella} E.,  et~al., 2016, preprint, \href
  {http://adsabs.harvard.edu/abs/2016arXiv161201526V} {} (\mn@eprint {arXiv}
  {1612.01526})

\bibitem[\protect\citeauthoryear{{Villegas} et~al.,}{{Villegas}
  et~al.}{2010}]{Villegas2010}
{Villegas} D.,  et~al., 2010, \mn@doi [\apj] {10.1088/0004-637X/717/2/603},
  \href {http://adsabs.harvard.edu/abs/2010ApJ...717..603V} {717, 603}

\bibitem[\protect\citeauthoryear{{Vogelsberger} et~al.,}{{Vogelsberger}
  et~al.}{2014}]{Vogelsberger2014}
{Vogelsberger} M.,  et~al., 2014, \mn@doi [\mnras] {10.1093/mnras/stu1536},
  \href {http://adsabs.harvard.edu/abs/2014MNRAS.444.1518V} {444, 1518}

\makeatother
\end{thebibliography}
%
\appendix
\section{Additional diagrams.}
\newpage
\begin{figure}
	\includegraphics[width=\columnwidth]{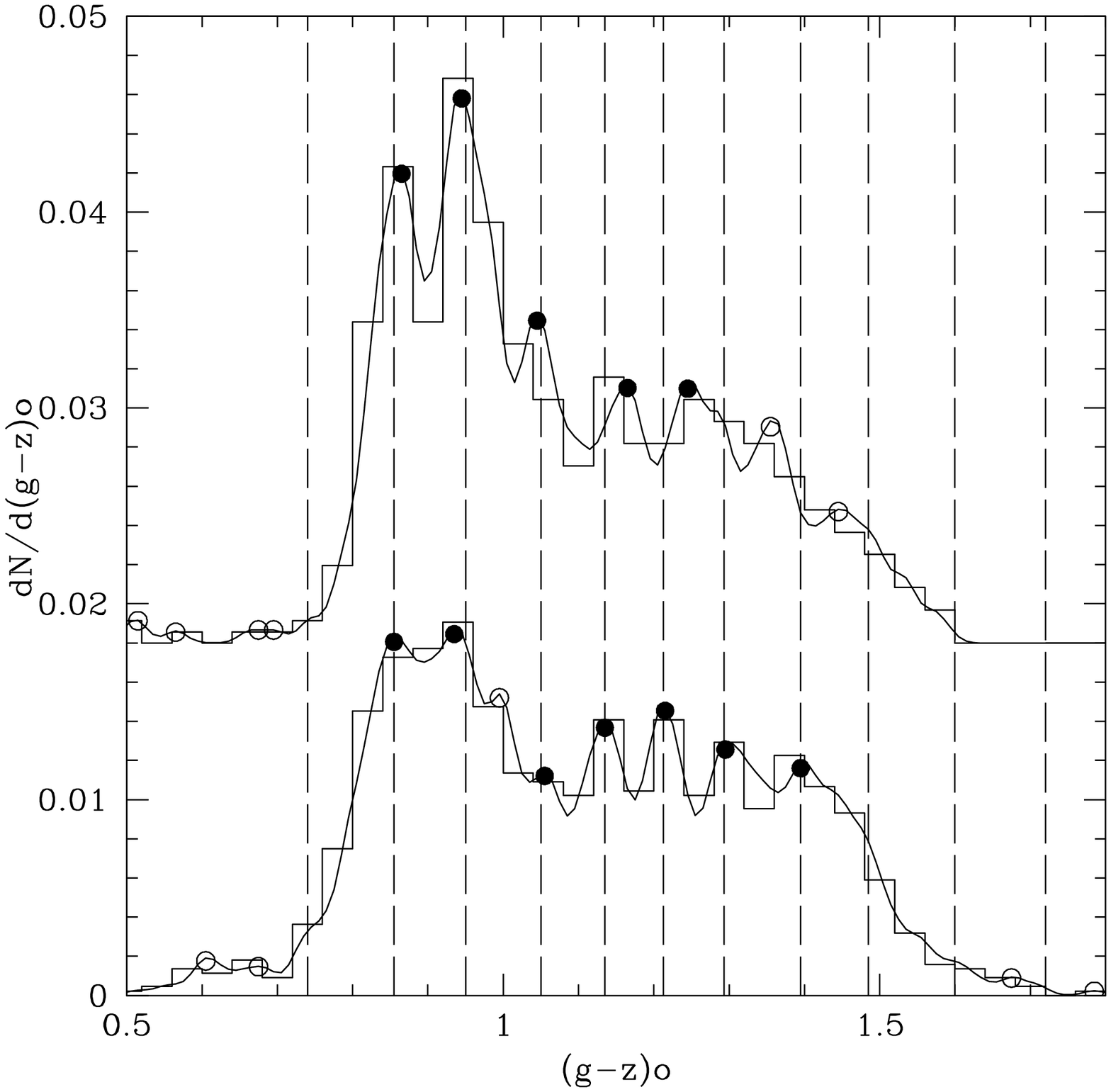}
    \caption{Discrete and smoothed (g-z)o colour distribution for $GCs$ in 13 Virgo galaxies with $M_{g}$
 between -20.2 and -19.2. The histogram have 0.04 mag bins while the smoothed distributions 
 correspond to a 0.015 mag Gaussian kernel. The upper histogram (arbitrarily shifted upwards) includes 437 $GCs$ with 
 apparent magnitude $g$ from 20 to 23. The lower histogram includes 1098 $GCs$ 
 with $g$ from 23.0 to 25.0.
}
    \label{fig:a1}
\end{figure}
\begin{figure}
	\includegraphics[width=\columnwidth]{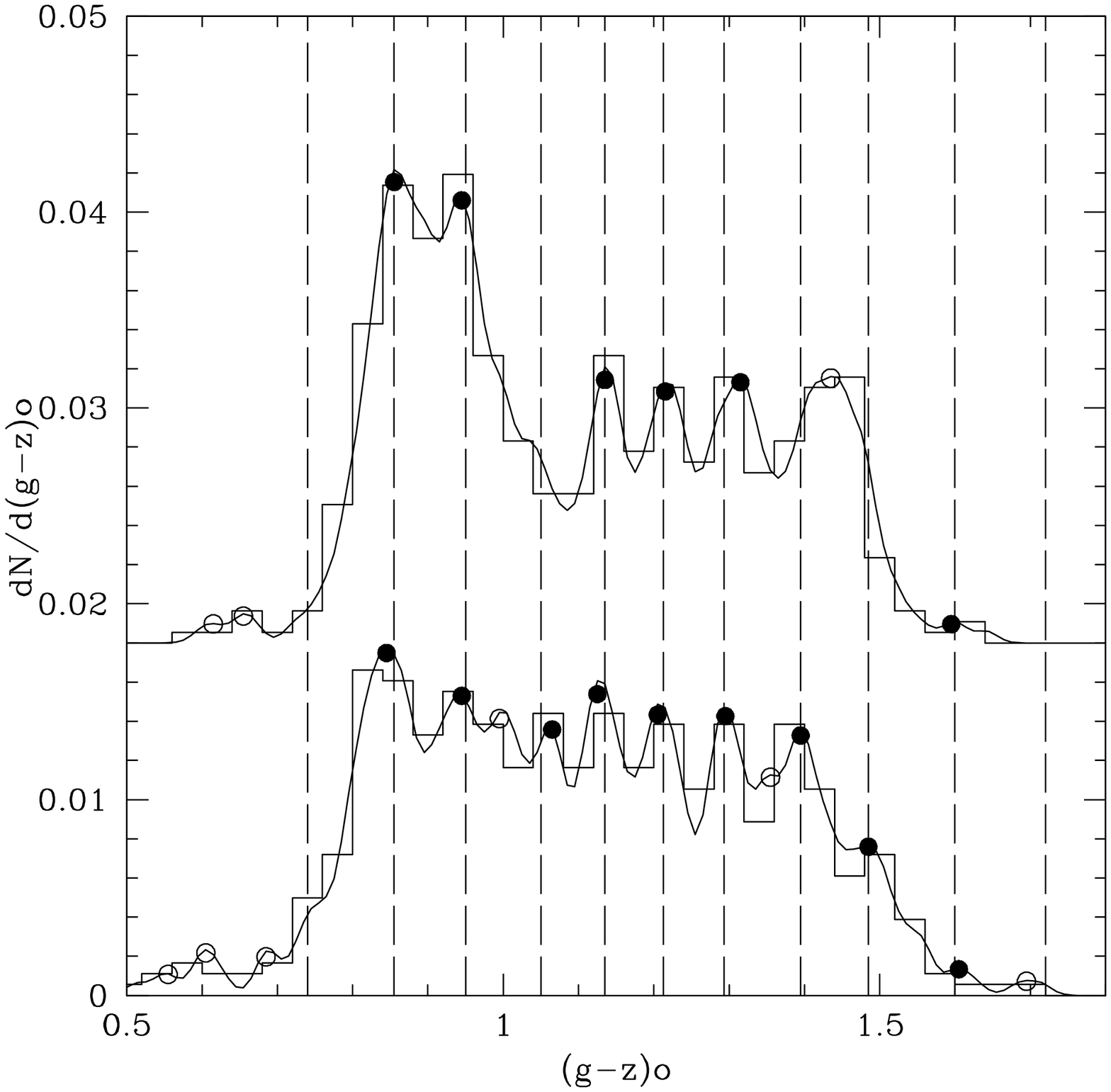}
    \caption{Discrete and smoothed (g-z)o colour distribution for $GCs$ in 13 Virgo galaxies with $M_{g}$
 between -20.2 and -19.2.  The upper histogram (arbitrarily shifted upwards) corresponds to $GCs$ with $g=$ 23.0 to 23.75 and 
 includes 458 clusters; the lower one corresponds to 450 $GCs$ with $g=$ 23.75 to 24.50. 
}
    \label{fig:a2}
\end{figure}
\begin{figure}
	\includegraphics[width=\columnwidth]{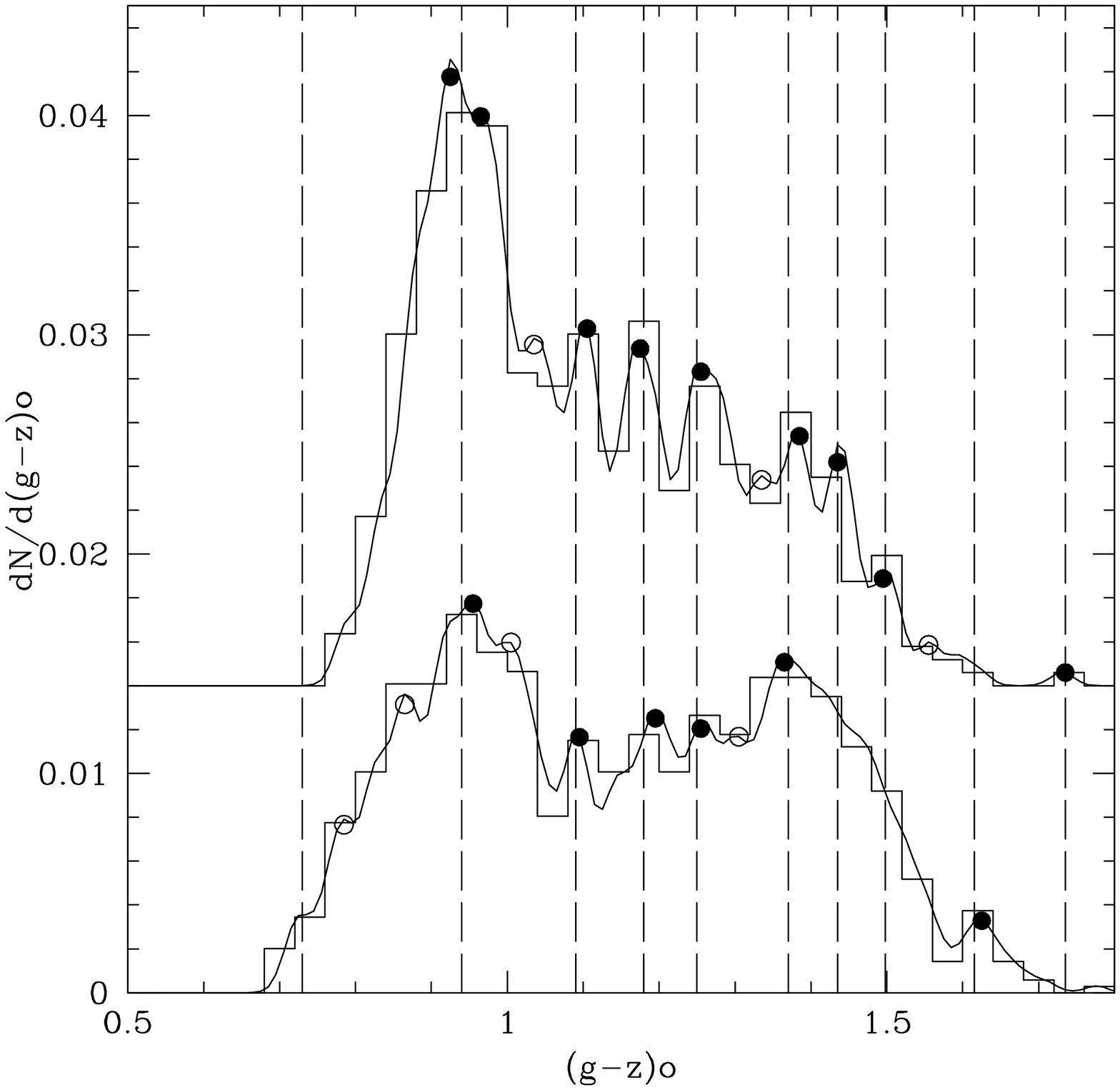}
    \caption{Discrete and smoothed (g-z)o colour distribution for $GCs$ in 7 Fornax galaxies with $M_{g}$
 between -20.2 and -19.2. The histograms have 0.04 mag bins while the smoothed distributions 
 corresponds to a 0.015 mag Gaussian kernel. The upper histogram (arbitrarily shifted upwards) includes 417 $GCs$ with 
 apparent magnitude $g$ from 20.0 to 23.4. The lower histogram corresponds to 852 $GCs$ with $g$
 from 23.4 to 25.0.
}
    \label{fig:a3}
\end{figure}
\begin{figure}
	\includegraphics[width=\columnwidth]{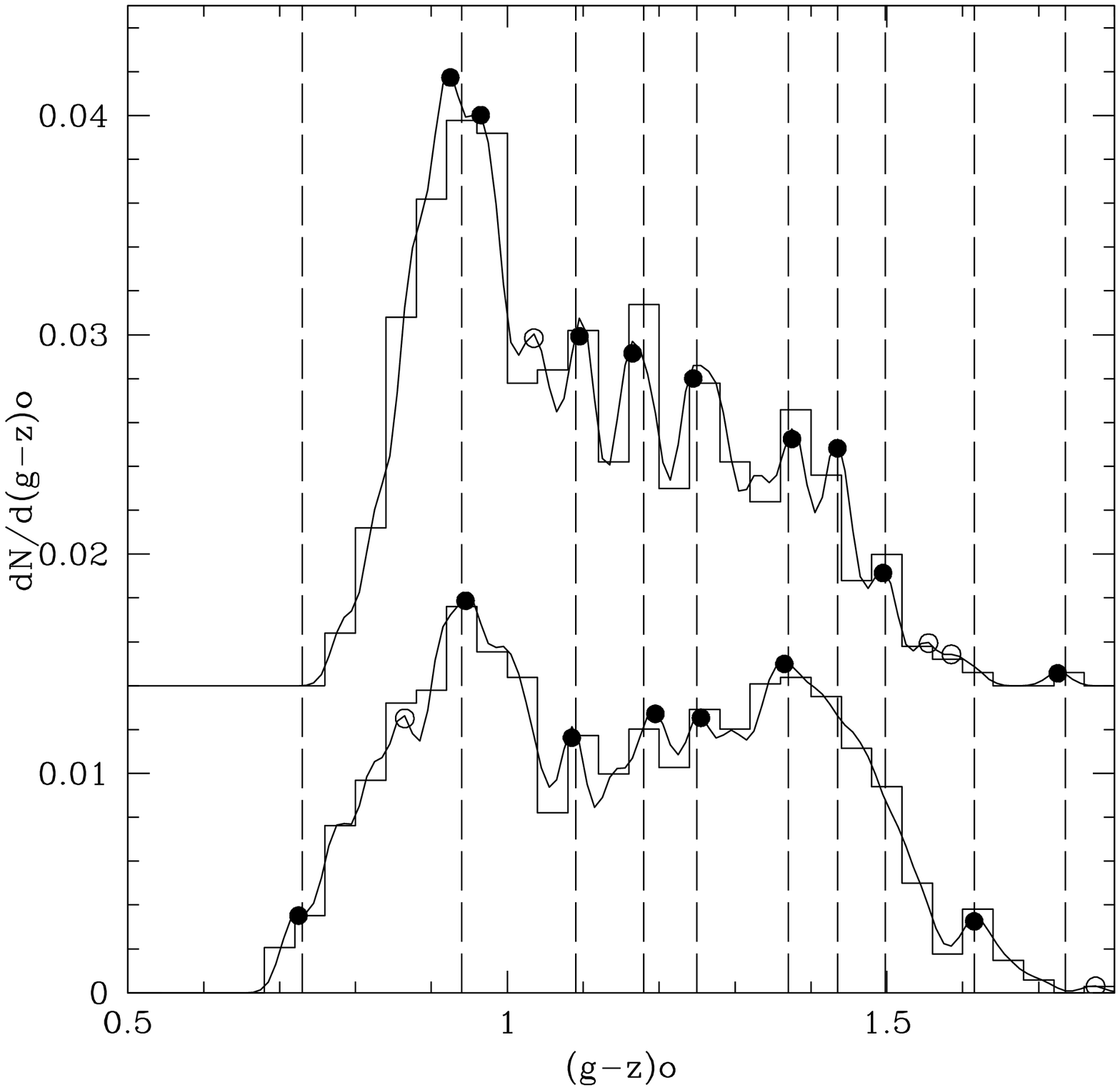}
    \caption{Discrete and smoothed $(g-z)o$ colour distribution for $GCs$ in 7  Fornax galaxies with $M_{g}$
 between -20.2 and -19.2.The upper histogram (arbitrarily shifted upwards) includes 417 $GCs$ with $g=$ 20.0 to 23.4 ; the lower
 one corresponds to 413 $GCs$ with  $g=$ 23.40 to 24.20.
}
    \label{fig:a4}
\end{figure}
\begin{figure}
	\includegraphics[width=\columnwidth]{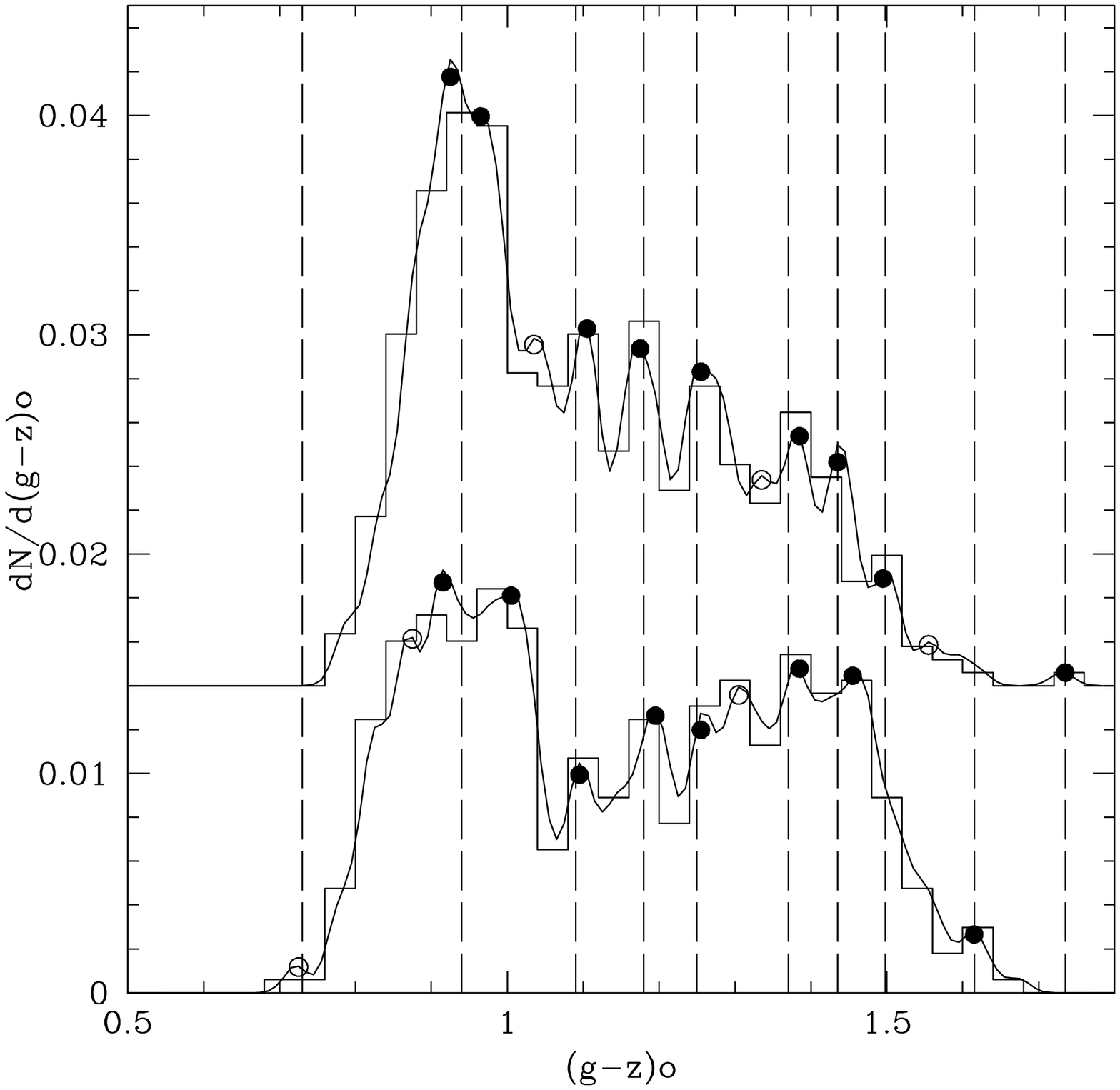}
    \caption{Discrete and smoothed $(g-z)o$ colour distribution for 522 $GCs$ candidates in an offset field in $NGC~4486$  and
 observed with Gemini-GMOS (upper histogram, arbitrarily shifted upwards), compared with that of $GCs$ in Virgo galaxies with 
 $M_{g}=$-20.2 to -19.2 (lower curve) and with the $TVP$ (dashed lines, see text). 
}
    \label{fig:a5}
\end{figure}
\begin{figure}
	\includegraphics[width=\columnwidth]{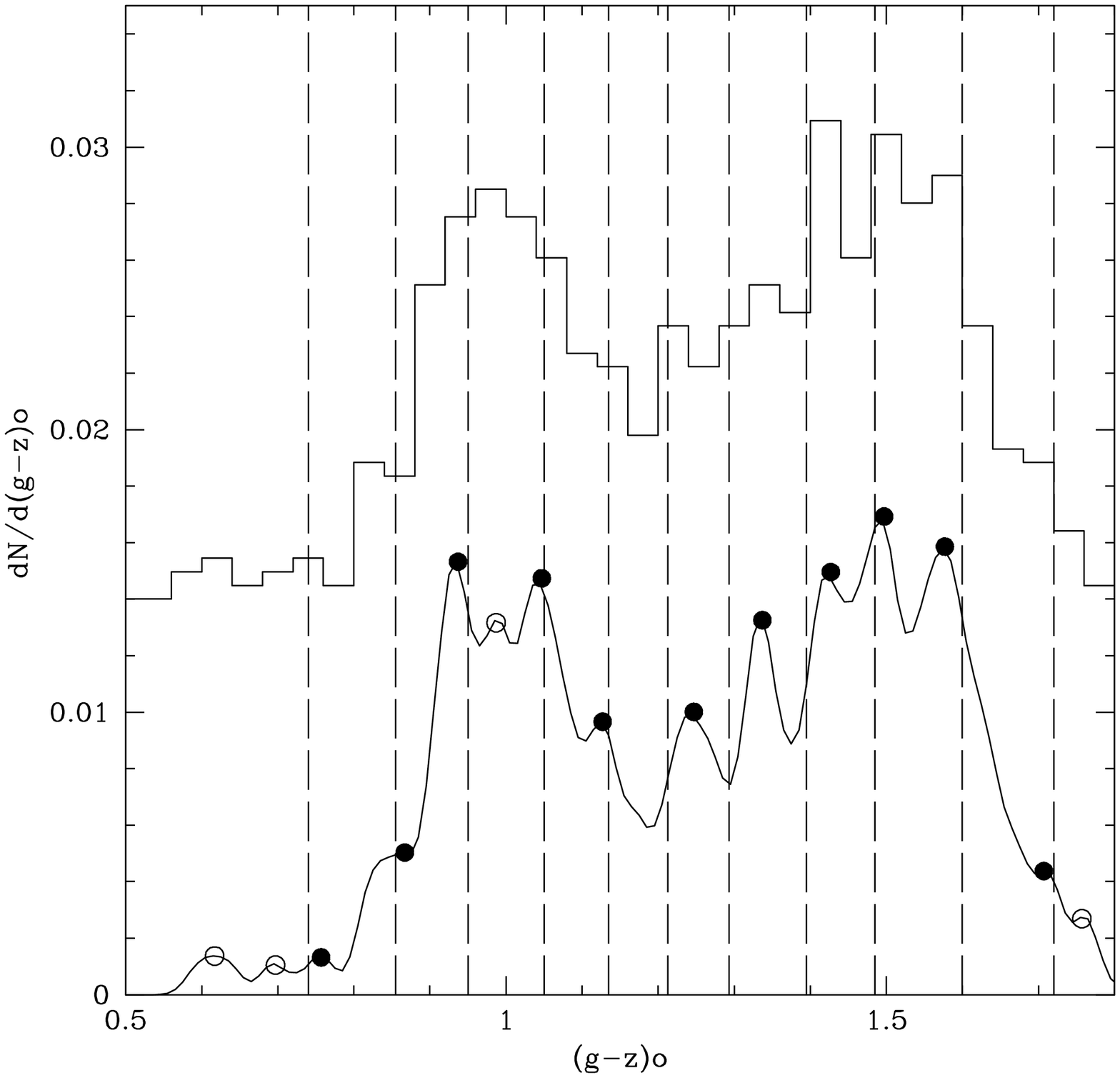}
    \caption{Discrete histogram (arbitrarily shifted upwards) and smoothed $(g-z)o$ colour distribution for 517 $GCs$ in $NGC~4472$.
The clusters have galactocentric radii between 20 and 110 $\arcsec$. This $GCs$ sample exhibits
the double blue peak seen in the $TVP$ and seven coincidences [4,5,6,7,8,9,10 and possibly 11] in the
domain of the intermediate and red $GCs$.
}
    \label{fig:a6}
\end{figure}
\begin{figure}
	\includegraphics[width=\columnwidth]{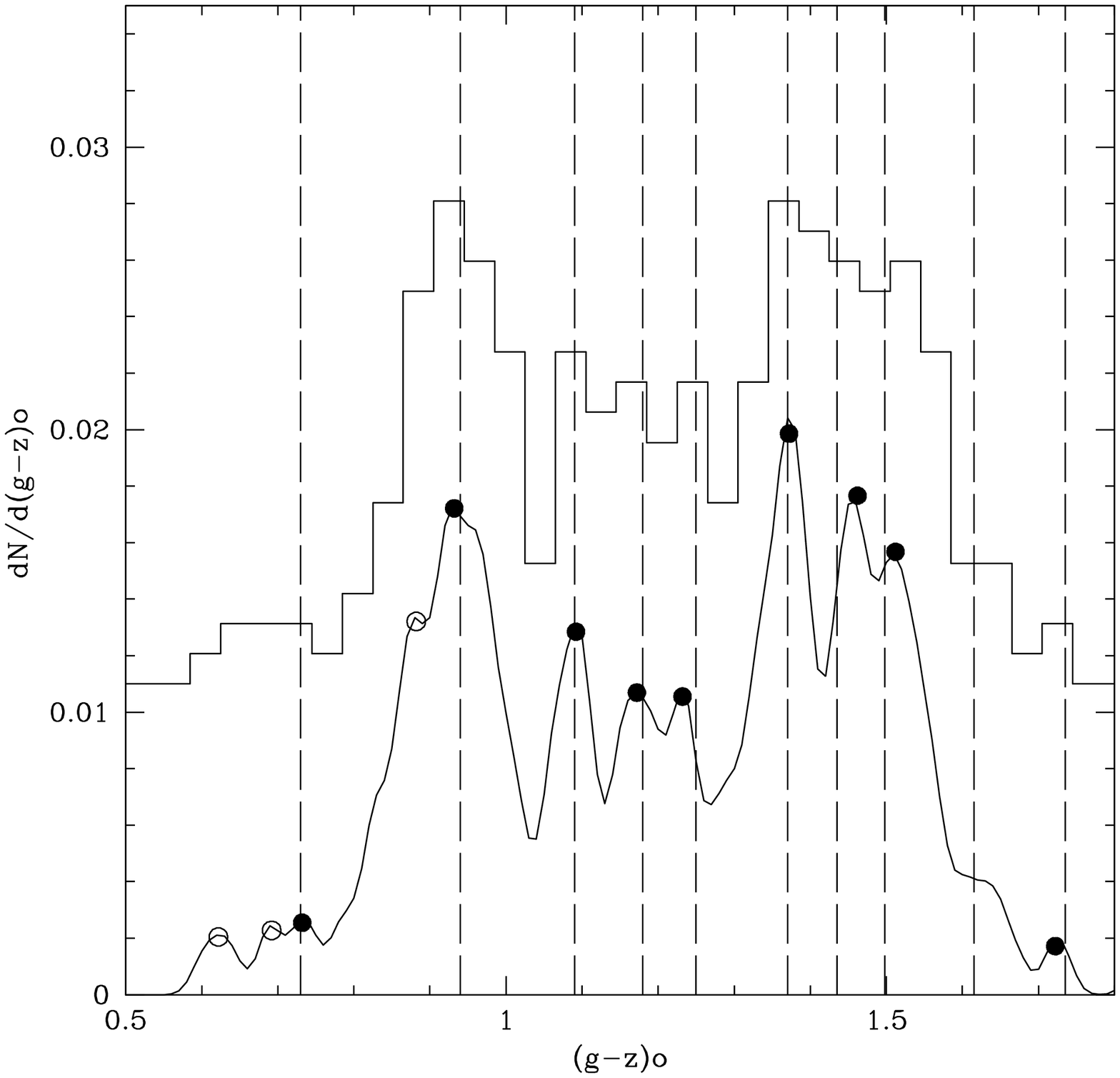}
    \caption{Discrete histogram (arbitrarily shifted upwards) and smoothed $(g-z)o$ colour distribution for $234$ $GCs$ in $NGC~1404$.
The clusters have galactocentric radii between 0 and 80 $\arcsec$.
}
    \label{fig:a7}
\end{figure}
\begin{figure}
	\includegraphics[width=\columnwidth]{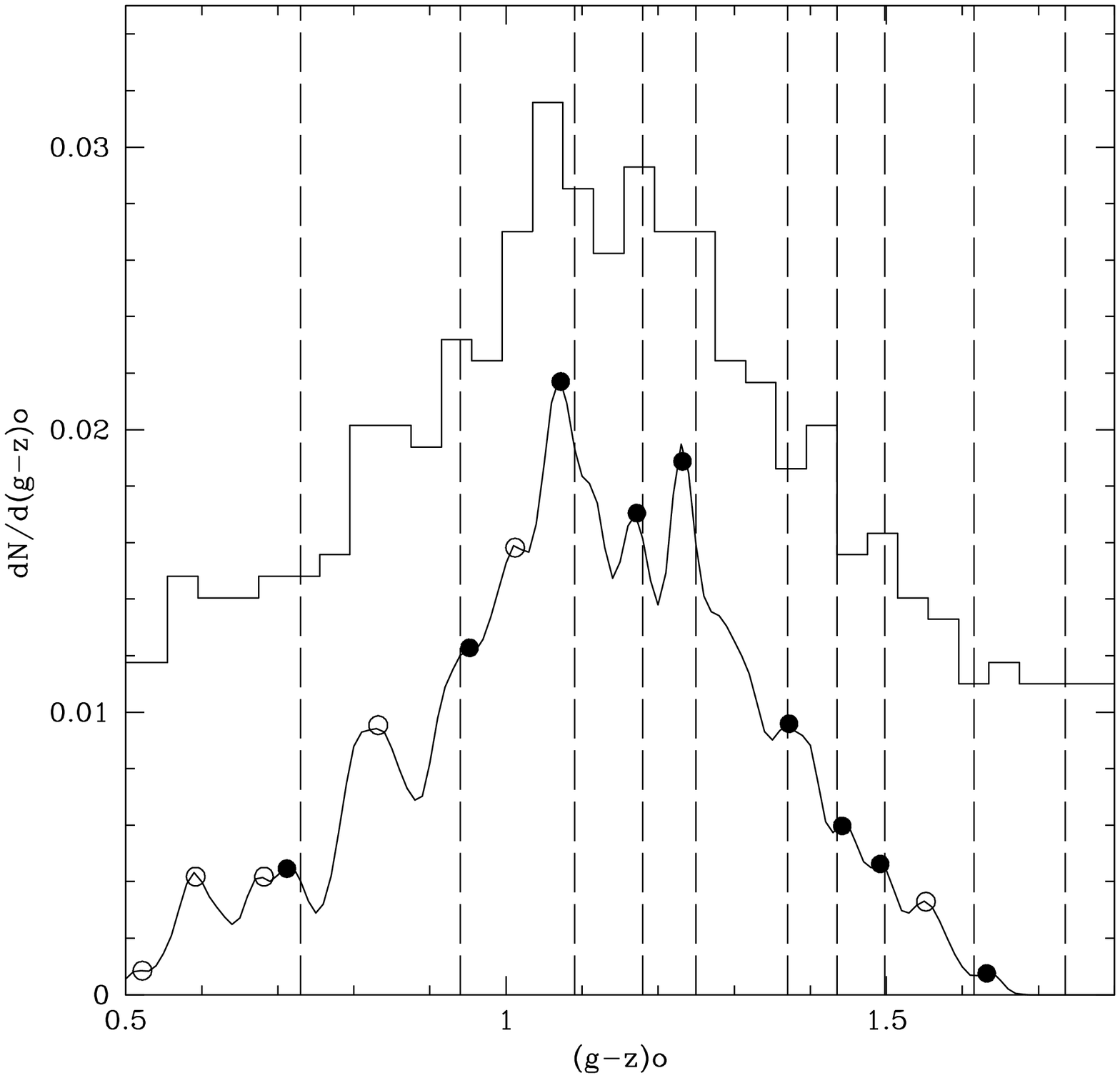}
    \caption{Discrete histogram (arbitrarily shifted upwards) and and smoothed $(g-z)o$ colour distribution for 367 $GCs$  with galactocentric
 radii from 40 to 120 $\arcsec$ in $NGC~1316$. Several colour peaks  on top of a broad $GCs$ colour distribution
 are coincident with those defined by the $TFP$. This colour distribution shows an extended blue tail arising in younger cluster populations (see text).
}
    \label{fig:a8}
\end{figure}
--------------------------------------------
\bsp	
\label{lastpage}
\end{document}